\documentclass[lettersize,journal]{IEEEtran}
\usepackage{amsmath,amsfonts}
\usepackage{algorithmic}
\usepackage{algorithm}
\usepackage{array}
\usepackage[caption=false,font=normalsize,labelfont=sf,textfont=sf]{subfig}
\usepackage{textcomp}
\usepackage{stfloats}
\usepackage{url}
\usepackage{verbatim}
\usepackage{graphicx}
\usepackage{cite}
\usepackage{color}
\usepackage{amssymb}
\usepackage{bm}
\usepackage{booktabs}
\usepackage{tabularx}
\begin{document}

\title{Secure Directional Modulation Enabled by Elevatable and Rotatable Antenna Array for Maritime Communications}

\author{Maolin Li, Haini Wu, Feng Shu, Yongpeng Wu, Cunhua Pan, Jianghzou Wang
	\thanks{(Corresponding author: Feng Shu.) }
		\thanks{Maolin Li and Haini Wu are with the School
		of Information and Communication Engineering, Hainan University, Haikou,
		570228, China. (e-mail: limaolin0302@163.com).}
	\thanks{Feng Shu is with the School of Information and Communication Engineering, Hainan University, Haikou 570228, China, and also with the School of Electronic and Optical Engineering, Nanjing University of Science and Technology, Nanjing 210094, China. (e-mail: shufeng0101@163.com).}
	\thanks{Yongpeng Wu is with the Shanghai Key Laboratory of Navigation and Location Based Services, Shanghai Jiao Tong University, Minhang, Shanghai, 200240, China (e-mail: yongpeng.wu2016@gmail.com).}
		\thanks{Cunhua Pan is with the National Mobile Communications Research Laboratory, Southeast University, China. (e-mail: cpan@seu.edu.cn).}
		\thanks{Jiangzhou Wang is with the School of Information Science and Engineering, Southeast University, Nanjing 211111, China (e-mail: j.z.wang@seu.edu.cn).}
}



\maketitle

\begin{abstract}
To address the dual challenges of limited transmission distance and wireless communication security in maritime communications, a directional modulation network enhanced by an elevatable and rotatable antenna array is investigated in this paper. Specifically, the base station, whose array and antenna elements can be rotated and whose antenna height can be adjusted, transmits information to legitimate users while resisting eavesdropping. Moreover, transceiver hardware impairments are taken into account. Accordingly, a worst‑case secrecy rate maximization problem is formulated under the joint optimization of precoding vectors, array orientation, antenna orientations, and array height. To solve this non‑convex optimization challenge, an alternating optimization scheme and a Transformer‑based learning scheme are proposed for the single‑user and multi‑user scenarios, respectively. Simulation results validate the effectiveness of the proposed algorithms. Security performance gains can be obtained after optimizing the array height, array orientation, and antenna orientations. In addition, compared with the isotropic antenna benchmark, the Transformer‑based learning scheme can achieve approximately 90\% power savings, 87\% antenna savings, and serve more legitimate users.
\end{abstract}


\section{Introduction}
Maritime wireless communications are becoming increasingly important for transportation, monitoring, rescue, shipping, and resource exploration~\cite{Yang2024,Liu2026a,Chen2026}. Compared with terrestrial communication systems, maritime communication networks are typically required to provide reliable wireless connectivity over longer propagation distances. However, due to sparse infrastructure, the Earth’s curvature, and severe propagation attenuation, it is difficult to simultaneously guarantee communication distance and quality~\cite{Sun2026}. In addition, the openness and broadcast nature of maritime wireless channels make confidential information vulnerable to unauthorized interception~\cite{Xu2023}. Therefore, extending the effective communication range while enhancing wireless transmission security is a significant challenge for future maritime communication systems.

Physical layer security provides an effective means of protecting confidential transmissions by exploiting the disparity between legitimate and eavesdropping channels~\cite{Wu2018,Yu2026,Zhao2025}. Directional modulation can concentrate useful information toward the legitimate receiver while suppressing information leakage in undesired spatial directions~\cite{Qiu2024,Tao2025,Shu2024}. However, the performance of conventional directional modulation relies on fixed antenna arrays. When the spatial characteristics of the legitimate user and the eavesdropper differ only slightly, conventional beamforming schemes struggle to simultaneously guarantee user quality of service and communication security requirements. In addition, in maritime scenarios, the effective communication range is further affected by the antenna height, since the sea surface and the Earth’s curvature may limit the line‑of‑sight propagation range. To address these challenges, additional physical degrees of freedom are introduced into the antenna architecture for investigation.

The elevation and rotation capabilities of antenna arrays provide a promising solution to the above problems~\cite{Pi2026, Jiang2026,Shao2025b}. By adjusting the antenna height, the visible communication distance can be extended. By rotating the entire array, better communication conditions can be created for legitimate users. In addition, when each antenna element is independently rotatable and directional, the directional gains toward the legitimate user and the eavesdropper can be controlled. Array rotation, array displacement, and antenna rotation have been investigated as means to provide additional spatial degrees of freedom~\cite{Zhu2026}. According to their physical characteristics and implementation methods, various intelligent architectures have been widely studied, including reconfigurable intelligent surfaces~\cite{Yan2020,Huang2026,Zheng2024}, fluid antennas~\cite{Yuan2026,Xu2026}, movable antennas~\cite{Zhu2026a,Li2026c}, rotatable antennas~\cite{Peng2026,Zhang2026}, and pinching antennas~\cite{Li2026b,Ding2025}. However, the above technologies generally involve small‑scale movement of antenna elements, resulting in limited achievable performance gains and making them unable to handle long‑distance, high‑security maritime communications with sparsely distributed users.
In \cite{Shao2025} and \cite{Shao2025a}, six‑dimensional (6D) movable antenna architectures under discrete and continuous positions and rotations were investigated, respectively, and an iterative search method was employed to optimize all array positions. Significant sum‑rate gains were achieved compared with fixed‑position antenna arrays. In \cite{Liu2026, Wang2026}, a UAV equipped with a reconfigurable intelligent surface was considered as a 6D architecture to address the angle‑dependent minimum receive signal-to-interference-plus-noise ratio (SINR) maximization problem. Considering antenna kinematic constraints and system delays, a popular optimization scheme was proposed in \cite{Cheng2026} to maximize the total spectral efficiency. In addition, the investigation in \cite{Shao2026} illustrated that a reinforcement‑learning‑based 6D antenna system outperforms fixed‑position and fluid antenna systems~\cite{Han2025} in the near-field. To balance hardware cost and system performance, rotatable antennas are a promising solution~\cite{Zhu2026b,Li2026d}. Through electronic or mechanical means, the antenna boresight can be flexibly adjusted, thereby exploiting the potential degrees of freedom to improve channel conditions and enhance system performance~\cite{Zheng2026b}. In \cite{Zheng2026a}, a two‑stage algorithm was proposed to jointly optimize beamforming and the antenna's maximum radiation direction, achieving a higher minimum signal‑to‑interference‑plus‑noise ratio for users compared with conventional schemes. In addition, the advantages of rotatable antennas in enhancing sensing resolution and range are expected to become a key technology for future intelligent communications~\cite{Zheng2026}.

However, the above studies mainly consider terrestrial communication scenarios. Research on the improvement of physical‑layer security performance through antenna mobility in maritime communication scenarios is still in its infancy. In addition, in practical communications, transceiver hardware impairments are unavoidable due to the non‑ideal characteristics of RF devices, such as power amplifier nonlinearity, phase noise, and I/Q imbalance~\cite{He2026,Li2026,Zhao2026}. The resulting additional distortion further affects the reception performance of both legitimate users and eavesdroppers. Therefore, it is necessary to take hardware impairments into account in the design of secure maritime communications. In \cite{Li2026a}, a movable antenna‑aided directional modulation system was investigated with hardware impairments taken into account. However, since the array is fixed, it is difficult to address the security problem for users located in the sidelobe directions.

The main contributions of this paper are summarized as follows.
\begin{itemize}
	\item We propose an elevatable and rotatable antenna array architecture for maritime communications, which has lower implementation complexity than 6D movable antennas. Hardware impairments are taken into account to better reflect practical conditions. A worst‑case secrecy rate (SR) maximization scheme is proposed through the joint optimization of array height, array orientation, antenna orientation, and precoding vectors.
	\item For the single‑user scenario, an iterative optimization algorithm is proposed to solve the non‑convex optimization problem. The system performance variation caused by the array height is derived to provide a solution for height optimization. For the multi‑user scenario, a Transformer‑based learning scheme is proposed, aiming to improve system security performance by learning the correlations among variables.
	\item Simulation results validate the effectiveness of the proposed schemes. It is worth noting that, due to the high path loss in maritime communications, the security performance gain brought by increasing the array height is limited. Array and antenna rotations can achieve higher SR values. In addition, compared with the isotropic antenna benchmark, significant power savings and antenna savings can be achieved, and more users can be served at the same security level.
\end{itemize}

\subsection{Organization}
The remainder of this paper is organized as follows. Sec.~II presents the system model and formulates the worst-case SR maximization problem. Sec.~III investigates the single-user scenario and develops the proposed alternating optimization scheme. Sec.~IV presents the Transformer-based multi-user optimization scheme. Sec.~V provides the simulation results. Finally, Section~VI concludes the paper.


\section{System Model and Problem Formulation}\label{sec:2}
\subsection{System Model}
\begin{figure}
	\centering
	\includegraphics[width=0.44\textwidth, trim = 110 120 90 90,clip]{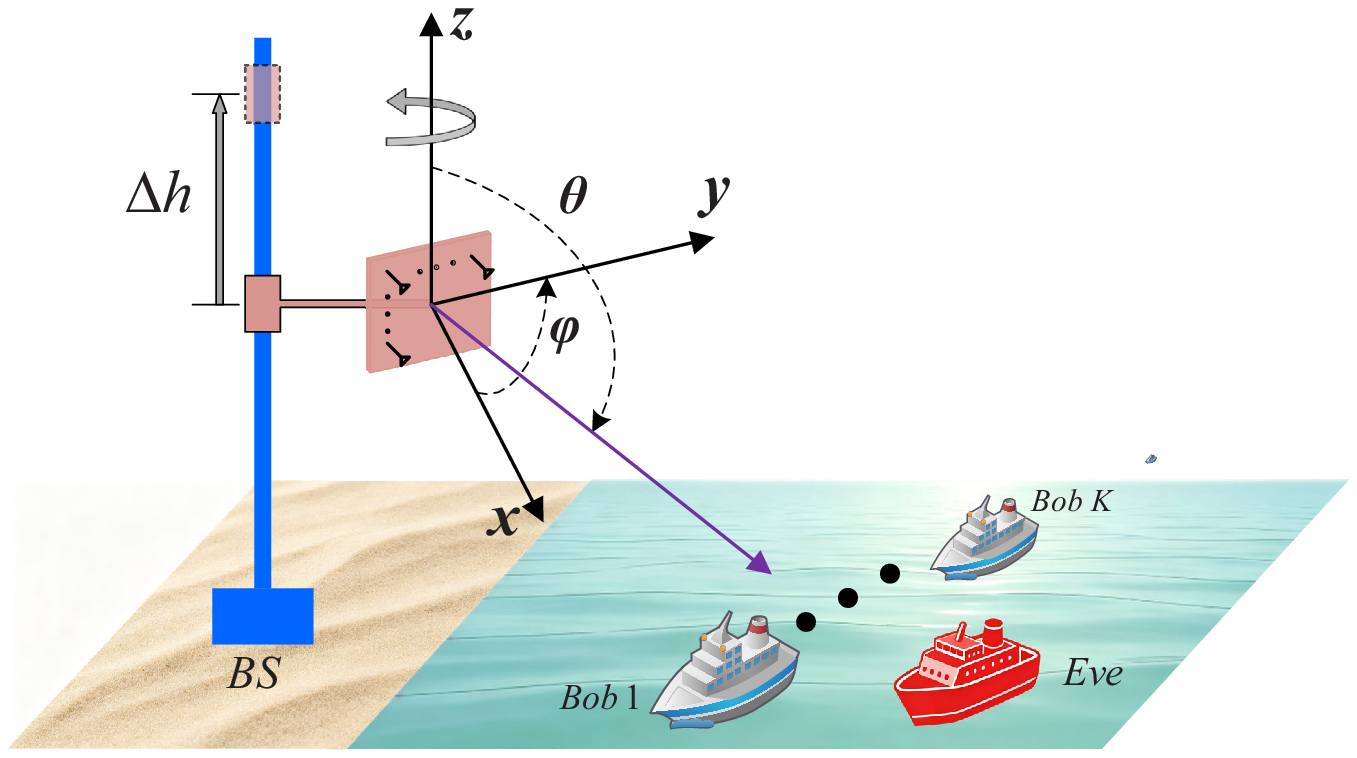}\\
	\caption{Illustration of an elevatable and rotatable antenna array-enabled directional modulation network.}\label{fig:1}
\end{figure}
As illustrated in Fig. \ref{fig:1}, we consider a secure maritime communication system enhanced by an elevatable and rotatable antenna array. Specifically, the BS is equipped with an $N$-element planar antenna array with fixed relative element positions. The array elements are directional rotatable antennas. The numbers of antennas in the horizontal and vertical directions are $N_y$ and $N_z$, respectively. The height and orientation of the array can be adjusted, where the orientation is determined by both the azimuth and elevation angles of the array. $K$ legitimate users (Bobs) and one eavesdropper (Eve) are considered, each equipped with a single antenna. Assume that the initial height of the array is $H_0$, and that its adjustable height increment is denoted by $\Delta h$. We establish a three-dimensional Cartesian coordinate system with the initial array center as the origin. The allowable rotation ranges of $\phi$ and $\theta$ are $[0, \phi_{\max}]$ and $[0,\theta_{\max}]$, respectively. $\phi_{\max}$ and $\theta_{\max}$ denote the maximum adjustable angles, respectively. The initial coordinates of the array elements can be expressed as  
\begin{align}
				r_{n,y}&=(\left\lfloor\frac{\left(n-1\right)}{N_z}\right\rfloor-\frac{N_y-1}{2})d_0,\\
				r_{n,z}&=(n-1-\left\lfloor\frac{\left(n-1\right)}{N_z}\right\rfloor N_z-\frac{N_z-1}{2})d_0,
\end{align}
where $n\in\mathcal{N}=\{1,2,\ldots,N\}$ and $d_0$ denote the antenna index and the minimum spacing between antennas, respectively. 
Then, the initial position of the 
$n$-th antenna can be expressed as $\mathbf{p}_n=[0,r_{n,y},r_{n,z}]^T$.
The initial position of the $k$-th Bob can be expressed as $\mathbf{b}_k=[x_k,y_k,h_{R,k}-H_0]^T$, $k\in\mathcal{K}=\{1,2,\ldots,K\}$. The initial position of the Eve can be expressed as $\mathbf{e}=[x_e,y_e,h_{R,e}-H_0]^T$. $h_{R,k}$ and $h_{R,e}$ denote the receive antenna heights of the $k$-th Bob and Eve, respectively. After rotation, the position of the $n$-th antenna can be expressed as 
\begin{align}\label{eq6}
&\mathbf{p}_n^{\prime}(\theta,\phi)  = \mathbf{R}(\theta,\phi)\mathbf{p}_n,
\end{align}
where
\begin{align}
	\mathbf{R}(\theta, \phi) 
	&=
	\begin{bmatrix}
		\cos\theta\cos\phi & -\cos\theta\sin\phi & \sin\theta \\
		\sin\phi & \cos\phi & 0 \\
		-\sin\theta\cos\phi & \sin\theta\sin\phi & \cos\theta
	\end{bmatrix}.
	\label{eq:rotation}
\end{align}
After increasing the height by $\Delta h$, the position coordinates of the $n$-th antenna can be expressed as 
\begin{align}\label{eq5}
	&\mathbf{p}_n^{\prime\prime}(\theta,\phi,\Delta h)  = \mathbf{p}_n^{\prime}(\theta,\phi) + [0,0,\Delta h]^T.
\end{align}

 In communication systems, due to the complexity of radio frequency chains and manufacturing imperfections, hardware impairments are inevitable. The distortion noise caused by hardware impairments (HWIs) is proportional to the transmit power of each antenna. Thus, the transmitted signal can be expressed as 
 \begin{align}\tilde{\mathbf{x}}
 	&=\sum_{k=1}^{K} (\sqrt{\frac{\beta P_0}{K}} \mathbf{w}_k s_k + \sqrt{\frac{(1-\beta) P_0}{K}} \mathbf{v}_k z_k) + \bm{\eta}_t,\end{align}
 where $\beta$, $s_k$, and $z_k$ denote the power allocation factor, the confidential symbol for the $k$-th Bob, and the artificial noise symbol, respectively. $\mathbf{w}_k\in\mathbb{C}^{N\times 1}$ and $\mathbf{v}_k\in\mathbb{C}^{N\times 1}$ denote the precoding vectors corresponding to $s_k$ and $z_k$, respectively, $\|\mathbf{w}_k\|^2=1$ and $\|\mathbf{v}_k\|^2=1$. For clarity, $\mathbf{w}_k$ and $\mathbf{v}_k$ are defined as the information precoding vector and the artificial noise precoding vector, respectively. $\bm{\eta}_t\sim\mathcal{CN}(0,{\frac{\kappa_t P_0}{K}}( \beta\sum_{k=1}^{K} (\mathbf{w}_k\mathbf{w}_k^H)\odot\mathbf{I}_N +   (1-\beta)\sum_{k=1}^{K}(\mathbf{v}_k \mathbf{v}_k^H) \odot\mathbf{I}_N))$ denotes the hardware impairments corresponding to the BS, where $\kappa_t$ denotes the hardware impairment coefficient. 
 
 Let the channel from the base station to the $k$-th Bob and the channel to Eve be denoted as $\mathbf{h}_k$ and $\mathbf{h}_e$, respectively. Then, the signal received by the $k$-th Bob can be expressed as
  \begin{align}
  	\tilde{y}_k 
  	&= \underbrace{\sqrt{\frac{\beta P_0}{K}} \mathbf{h}_k^H\mathbf{w}_k s_k}_{\mathrm{Desired\ signal}} + \underbrace{\sum_{i=1,i\neq k}^{K}\sqrt{\frac{\beta P_0}{K}} \mathbf{h}_k^H\mathbf{w}_i s_i}_{\mathrm{Multi-user\ interference}} \notag\\&\quad+ \underbrace{\sum_{i=1}^{K}  \sqrt{\frac{(1-\beta) P_0}{K}} \mathbf{h}_k^H\mathbf{v}_iz_i }_{\mathrm{AN\ interference}} + \underbrace{\mathbf{h}_k^H\bm{\eta}_t+ {\eta}_r}_{\mathrm{Hardware\ impairments}} + n_k,
  \end{align}
 where $\eta_r\sim\mathcal{CN}(0,\kappa_r\mathbb{E}|y_k|^2)$ denotes the hardware impairments corresponding to the Bob, where $\kappa_r$ denotes the hardware impairment coefficient. We assume that the Eve is equipped with a perfect receiver. This assumption ensures that the information leakage caused by underestimating the Eve's reception capability is avoided. Accordingly, the signal received at Eve for decoding the confidential symbol intended for the \(k\)-th Bob can be expressed as
   \begin{align}
 	e_k &= \underbrace{\sqrt{\frac{\beta P_0}{K}} \mathbf{h}_e^H\mathbf{w}_k s_k}_{\mathrm{Desired\ signal}} + \underbrace{\sum_{i=1,i\neq k}^{K}\sqrt{\frac{\beta P_0}{K}} \mathbf{h}_e^H\mathbf{w}_i s_i}_{\mathrm{Multi-user\ interference}} \notag\\&\quad+ \underbrace{\sum_{i=1}^{K}  \sqrt{\frac{(1-\beta) P_0}{K}} \mathbf{h}_e^H\mathbf{v}_i z_i }_{\mathrm{AN\ interference}} + \underbrace{\mathbf{h}_e^H\bm{\eta}_t}_{\mathrm{Hardware\ impairments}}+ n_{e,k}.
 \end{align}
 $n_k\sim\mathcal{CN}(0,\sigma^2)$ and $n_{e,k}\sim\mathcal{CN}(0,\sigma^2)$ denote the noise experienced by the 
$k$-th Bob and Eve, respectively. Correspondingly, the achievable transmission rates for the $k$-th Bob and Eve are 
   \begin{align}
	R_k &= \log_2(1+\frac{\beta |\mathbf{h}_k^H\mathbf{w}_k|^2}{\sum_{i=1}^{3}a_{k,i}+
		(1+\kappa_r)K\sigma^2/P_0}),\label{9}\\
	R_{E,k} &= \log_2(1+\frac{\beta |\mathbf{h}_e^H\mathbf{w}_k|^2}{a_{k,4}+a_{k,5}+K\sigma^2/P_0}).
\end{align}
where
   \begin{align}
   	a_{k,1} &= (1+\kappa_r)\kappa_t\beta \mathbf{h}_k^H(\sum_{i=1}^{K}(\mathbf{w}_i\mathbf{w}_i^H)\odot\mathbf{I}_N)\mathbf{h}_k\notag\\&\quad+(1+\kappa_r)\kappa_t(1-\beta)\mathbf{h}_k^H(\sum_{i=1}^{K}(\mathbf{v}_i\mathbf{v}_i^H)\odot\mathbf{I}_N)\mathbf{h}_k,\\a_{k,2} &= 
   	(1+\kappa_r)\beta \sum_{i=1,i\neq k}^{K} |\mathbf{h}_k^H\mathbf{w}_i|^2+\kappa_r\beta|\mathbf{h}_k^H\mathbf{w}_k|^2,\\
   	a_{k,3} &= 
   	(1+\kappa_r)(1-\beta) \sum_{i=1}^{K} |\mathbf{h}_k^H\mathbf{v}_i|^2,\\
   	a_{k,4} &= \beta\sum_{i=1,i\neq k}^{K}| \mathbf{h}_e^H\mathbf{w}_i|^2+(1-\beta)\sum_{i=1}^{K}    |\mathbf{h}_e^H\mathbf{v}_i |^2,\\
   	a_{k,5} &= \kappa_t\beta \mathbf{h}_e^H(\sum_{i=1}^{K}(\mathbf{w}_i\mathbf{w}_i^H)\odot\mathbf{I}_N)\mathbf{h}_e\notag\\&\quad+\kappa_t(1-\beta)\mathbf{h}_e^H(\sum_{i=1}^{K}(\mathbf{v}_i\mathbf{v}_i^H)\odot\mathbf{I}_N)\mathbf{h}_e.
\end{align}
 Then, the worst‑case SR can be expressed as
    \begin{align}\label{16}
 	R_s &= \min_k\{[R_k-R_{E,k}]^+\}.
 \end{align}

\subsection{Channel Model}
To clearly characterize the channel variations caused by antenna height, we adopt the maritime communication channel model from~\cite{He2022}. The directional gain corresponding to the $n$-th rotatable antenna and the 
$x$-th ($x\in\{1,2,\ldots,K,e\}$) receiving user can be expressed as 
\begin{equation}
	G_{n,x}(\epsilon) =
	\begin{cases} 
		G_0 \cos^{2p}(\epsilon_{n,x}), & \epsilon_{n,x} \in \left[0, \frac{\pi}{2}\right), \\
		0, & \text{otherwise},
	\end{cases}
\end{equation}
where $p\geq0$ and $G_0=2(2p+1)$ denote the directivity factor and the maximum gain in the boresight direction, respectively. $\epsilon_{n,x}$ denotes the angle between the direction of the receiving user $x$ and the direction of the maximum antenna gain. The path loss of the link from the BS to the $x$-th receiving user can be denoted as
\begin{align}\label{18new}
\rho_x&=10^{-\frac{10\varsigma\log_{10}(\frac{d_x}{d_r})+\iota_x+P_d}{10}} = (\frac{d_x}{d_r})^{-\varsigma}10^{-\frac{\iota_x+P_d}{10}},
\end{align}
where $d_r$ represents the reference distance. $d_k=\|\mathbf{b}_k-[0,0,\Delta h]^T\|_2$ and $d_e=\|\mathbf{e}-[0,0,\Delta h]^T\|_2$ denote the distances from the array center to the $k$-th user and Eve, respectively. $\varsigma$ is the path loss exponent, which characterizes the attenuation rate with respect to the propagation distance. $\iota_x$ denotes the shadow fading in dB, and $P_d$ represents the reference path loss at the distance $d_r$. The receiver index $x\in\{1,\ldots,K,e\}$ corresponds to the
$k$-th Bob or Eve. Then, the channel from the BS to the $k$-th Bob can be denoted as $\mathbf{h}_k$, where
\begin{align}
	\!\mathbf{h}_k[n] &= (\rho_kG_{n,k})^{\frac{1}{2}}(\sqrt{\frac{\xi_R}{\xi_R + 1}} h_{\text{LoS}}^{n,k} + \sqrt{\frac{1}{\xi_R + 1}} h_{\text{NLoS}}^{n,k}).
\end{align}
The channel from the BS to the Eve can be denoted as $\mathbf{h}_e$, where
\begin{align}
	\mathbf{h}_e[n] &= (\rho_eG_{n,e})^{\frac{1}{2}}(\sqrt{\frac{\xi_R}{\xi_R + 1}} h_{\text{LoS}}^{n,e} + \sqrt{\frac{1}{\xi_R + 1}} h_{\text{NLoS}}^{n,e}).\label{18}
\end{align}
$\xi_R$ denotes the Ricean factor. 
Due to the earth’s curvature and evaporation ducts, the channel is typically modeled differently over different propagation ranges. Let $d_{n,x}\in\{d_{n,1},\ldots,d_{n,K},d_{n,e}\}$ denote the distance from the $n$-th antenna to receiver $x$. For the far‑field scenario, we have $d_{1,k}\approx d_{2,k}\approx \cdots\approx d_{N,k}$ and $d_{1,e}\approx d_{2,e}\approx \cdots\approx d_{N,e}$. The channel model in \eqref{18} is generally valid within the range $(0, d_{S,x})$, where 
$d_{S,n,x}$ is a distance parameter that can be expressed as
$d_{S,x} = \frac{4h_{R,x}(H_0+\Delta h)}{\lambda}.
$
 $\lambda$ denotes the the wavelength of the signal. $h_{\text{LoS}}^{n,k}$ and $h_{\text{LoS}}^{n,e}$ denote the LoS components corresponding to the $k$-th Bob and Eve, which can be expressed as 
\begin{align}
	h_{\text{LoS}}^{n,k} 
	=e^{j\frac{2\pi}{\lambda}\mathbf{u}_k^T(\theta_k,\phi_k,\Delta\theta_k)\mathbf{R}(\theta,\phi)\mathbf{p}_n },\\
	\!h_{\text{LoS}}^{n,e} 
	=e^{j\frac{2\pi}{\lambda}\mathbf{u}_e^T(\theta_e,\phi_e,\Delta\theta_e)\mathbf{R}(\theta,\phi)\mathbf{p}_n},
\end{align}
where \begin{align}&\mathbf{u}_k=\begin{bmatrix}
		\cos(\theta_k+\Delta\theta_k)\sin\phi_k\\\cos(\theta_k+\Delta\theta_k)\cos\phi_k\\\sin(\theta_k+\Delta\theta_k)
	\end{bmatrix},\\&\mathbf{u}_e=\begin{bmatrix}\cos(\theta_e+\Delta\theta_e)\sin\phi_e\\\cos(\theta_e+\Delta\theta_e)\cos\phi_e\\\sin(\theta_e+\Delta\theta_e)\end{bmatrix}.\end{align} 
Here, $\cos(\epsilon_{n,k})\triangleq \vec{\mathbf{p}}_n^T\vec{\mathbf{d}}_{n,k}$ represents the projection between $\vec{\mathbf{p}}_n$ and $\vec{\mathbf{d}}_{n,k}$, where 
\begin{align}&\vec{\mathbf{p}}_n=\begin{bmatrix}
		\cos\theta_{z,n}\sin\phi_{a,n}\ \cos\theta_{z,n}\cos\phi_{a,n}\ \sin\theta_{z,n}
	\end{bmatrix}^T\end{align}denotes the pointing vector, and \begin{align}&\vec{\mathbf{d}}_{n,k}=\frac{ \mathbf{b}_k-\mathbf{p}_n^{\prime\prime}(\theta,\phi,\Delta h)}{\| \mathbf{b}_k-\mathbf{p}_n^{\prime\prime}(\theta,\phi,\Delta h)\|}\end{align} 
denotes the direction vector from the $n$-th antenna to the $k$-th Bob. $\theta_{z,n}$ and $\phi_{a,n}$ denote the elevation and azimuth angles with respect to the local reference frame of the $n$-th antenna’s boresight rotation, respectively. Similarly, $\cos(\epsilon_{n,e})\triangleq \vec{\mathbf{p}}_n^T\vec{\mathbf{d}}_{n,e}$ represents the projection between $\vec{\mathbf{p}}_n$ and $\vec{\mathbf{d}}_{n,e}$, where $\vec{\mathbf{d}}_{n,e}=\frac{ \mathbf{e}-\mathbf{p}_n^{\prime\prime}(\theta,\phi,\Delta h)}{\| \mathbf{e}-\mathbf{p}_n^{\prime\prime}(\theta,\phi,\Delta h)\|}$.
$\theta_x$ and $\phi_x$ denote the elevation angle and azimuth angle corresponding to receiver $x$, respectively. $\Delta\theta_k$ and $\Delta\theta_e$ denote the angle variations associated with $\Delta h$, respectively. Through geometric derivation, we can obtain
\begin{align}
	&\theta_k + \Delta\theta_k = \operatorname{atan2} \left( h_{R,k} - H_0 - \Delta h, \sqrt{x_k^2 + y_k^2} \right),\\
	&\theta_e + \Delta\theta_e = \operatorname{atan2} \left( h_{R,e} - H_0 - \Delta h, \sqrt{x_e^2 + y_e^2} \right).
\end{align}
 The NLoS component is modeled as spatially independent circularly
 symmetric complex Gaussian fading, i.e.,
 \begin{align}
 	h_{\text{NLoS}}^{n,k}\sim\mathcal{CN}({0},1),\ h_{\text{NLoS}}^{n,e}\sim\mathcal{CN}({0},1),
 	\ \forall n\in\mathcal{N}, \forall k\in\mathcal{K}.
 \end{align}

\subsection{Problem Formulation}
We aim to maximize the worst‑case SR by jointly optimizing the precoding vectors for the $K$ Bobs, the $K$ AN precoding vectors, the power allocation factor, the rotation angles of the array, and the orientations of the $N$ rotatable antennas. Let $\mathcal{S}
\triangleq
\left\{
\{\mathbf{w}_k,\mathbf{v}_k\}_{k=1}^{K},
\beta,\Delta h,\theta,\phi,
\{\theta_{z,n},\phi_{a,n}\}_{n=1}^{N}
\right\}$ denote the set of all variables. Then, the optimization problem can be formulated as 
\begin{subequations}\label{P1}
	\begin{align}
		\mathrm{P1}:\quad
		\max_{\mathcal{S}}
		\quad&
		\min_{k\in\mathcal{K}}
		\left[
		R_k-R_{E,k}
		\right]^+
		\label{P1a}
		\\
		\mathrm{s.t.}\quad
		&R_k\geq R_{\min}, \quad \forall k\in\mathcal{K},
		\label{P1b}\\
		&\sum_{i=1}^{K}\mathbf{h}_k^H\mathbf{v}_i = 0, \quad \forall k\in\mathcal{K},
		\label{P1c}\\
		&\|\mathbf{w}_k\|^2=1,
		\quad \forall k\in\mathcal{K},
		\label{P1d}\\
		&\|\mathbf{v}_k\|^2=1,
		\quad \forall k\in\mathcal{K},
		\label{P1e}\\
		&0\leq\beta\leq1,
		\label{P1f}\\
		&0\leq\Delta h\leq\Delta h_{\max},
		\label{P1g}\\
		&0\leq\theta\leq\theta_{\max}, 0\leq\phi\leq\phi_{\max},
		\label{P1h}\\
		&0\leq\theta_{z,n}\leq\theta_{z,\max},
		\quad \forall n\in\mathcal{N},
		\label{P1i}
	\end{align}
\end{subequations}
Constraint \eqref{P1b} denotes the user quality-of-service (QoS) requirement, and $R_{\min}$ denotes the minimum transmission rate. Constraint \eqref{P1c} indicates that the artificial noise does not cause interference to the legitimate users. Constraints \eqref{P1d} and \eqref{P1e} impose the unit-norm requirements on the confidential-signal and AN precoding vectors, respectively. Constraint \eqref{P1f} restricts the power allocation factor. \eqref{P1g} and  \eqref{P1h} specify the feasible elevation and rotation ranges of the antenna array. Constraints \eqref{P1i} characterizes the feasible boresight orientation ranges of each rotatable antenna element.
\section{Single‑User Optimization Scheme}
\begin{figure}[t]
	\centering
	\subfloat[]{%
		\includegraphics[width=0.2\textwidth,
		trim=20 10 10 10,clip]{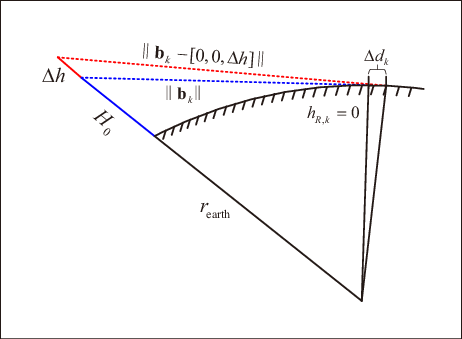}
		\label{fig:2-1}
	}
	\hfill
	\subfloat[]{%
		\includegraphics[width=0.25\textwidth,
		trim=10 10 10 10,clip]{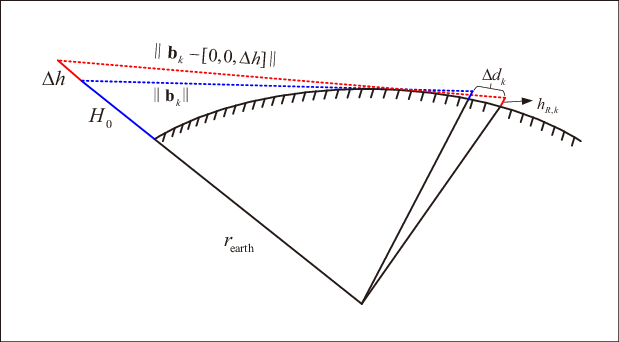}
		\label{fig:2-2}
	}
	\caption{Illustration of the maximum visible distance: (a) receiver antenna height is zero; (b) receiver antenna height is nonzero.}
	\label{fig:2}
\end{figure}
In this section, a single‑user scenario is considered, i.e., $K=1$. First, the relationship between the increase in antenna height and the propagation distance is derived using geometric principles. Second, the optimization problem P1 is solved in the single‑user scenario. For clarity, the algorithm proposed in this section is referred to as the iterative optimization (AO) algorithm.
\subsection{Communication Distance Setup}
 Due to the spherical shape of the Earth, the curved sea surface blocks the line of sight. As illustrated in Fig. \ref{fig:2-1}, $r_{\mathrm{earth}}$ denotes the earth’s radius, and $\Delta d_k$ denotes the change in the visible distance over the sea surface after the array height is increased by $\Delta h$, assuming that the receiver antenna height $h_{R,k}$ is zero. The maximum propagation distance is defined as the farthest distance visible from the array center in free space. According to the geometric relationship, the maximum separation distance $\Delta d_k$ of the receiver can be expressed as
\begin{align}
	\Delta d_k
	&=
	r_{\rm earth}
	\Bigg[
	\arccos\left(
	\frac{r_{\rm earth}}
	{r_{\rm earth}+H_0+\Delta h}
	\right)
	\notag\\
	&\qquad -
	\arccos\left(
	\frac{r_{\rm earth}}
	{r_{\rm earth}+H_0}
	\right)
	\Bigg].\label{34}
\end{align}
However, due to path loss and limited system resources, \eqref{34} is generally difficult to satisfy. For clarity, simplification and normalization operations are adopted. Under the assumptions of a single transmit–receive antenna and unit antenna gain, i.e., $N=1$ and $K=1$, the maximum separation distance is derived. According to \eqref{9} and \eqref{18new}, the transmission rate can be simplified as 
\begin{align}\label{35}
	R_1&=\log_2 (1+
	\frac{1}
	{(1+\kappa_r)\kappa_t+\kappa_r+(1+\kappa_r)\sigma^2/\tau_1}),
\end{align}
where
	$\tau_1=\beta P_0G_{1,1}(\frac{d_1}{d_r})^{-\varsigma}10^{-\frac{\iota_1+P_d}{10}}$.
 Then, the maximum communication distance under the minimum QoS requirement can be expressed as 
\begin{align}\label{37}
	d_1&\leq d_r\left[\frac{\beta P_0G_{1,1}10^{-\frac{\iota_1+P_d}{10}}({\frac{1}{2^{R_{\min}}-1 }- (1 + \kappa_r)\kappa_t - \kappa_r})}{{(1 + \kappa_r)\sigma^2}}\right]^{\frac{1}{\varsigma}}\notag\\&=d_{\max}.
\end{align}
Correspondingly, after the antenna height is increased by $\Delta h$, the change in the distance of the receiving point is 
\begin{align}
	\Delta d_1
	&=
	r_{\rm earth}
	\Bigg[
	\arccos\left(
	\frac{r_{\rm earth}^2+(r_{\rm earth}+H_0)^2-d_{\max}^2}
	{2r_{\rm earth}(r_{\rm earth}+H_0)}
	\right)
	\notag\\
	&-
	\arccos\left(
\frac{r_{\rm earth}^2+(r_{\rm earth}+H_0+\Delta h)^2-d_{\max}^2}
{2r_{\rm earth}(r_{\rm earth}+H_0+\Delta h)}
	\right)
	\Bigg].\label{38}
\end{align}
As illustrated in Fig. \ref{fig:2-2}, when the receiver antenna height $h_{R,1}$ is nonzero, the corresponding distance variation is 
\begin{align}
	&\Delta d_1=\notag\\
	&
	r_{\rm earth}
	\Bigg[
	\arccos\left(
	\frac{(r_{\rm earth}+h_{R,1})^2+(r_{\rm earth}+H_0)^2-d_{\max}^2}
	{2(r_{\rm earth}+h_{R,1})(r_{\rm earth}+H_0)}
	\right)
	\notag\\
	&-
	\arccos\left(
	\frac{(r_{\rm earth}\!\!+\!h_{R,1})^2+(r_{\rm earth}\!+\!H_0\!+\!\Delta h)^2-d_{\max}^2}
	{2(r_{\rm earth}+h_{R,1})(r_{\rm earth}+H_0+\Delta h)}
	\right)
	\Bigg].\label{39}
\end{align}

According to \eqref{37}, under the predicted channel conditions, the signal strength at a distant eavesdropper can be reduced by lowering the transmit power and antenna gain. However, when the eavesdropper is closer to the base station, its channel quality may be superior to that of the legitimate users. Therefore, joint optimization of beamforming, antenna orientations, array pose, and power allocation is required to enhance system security. In addition, how to exploit the degree of freedom introduced by the elevatable antenna to enhance system security within the effective transmission range remains an important challenge. For tractability, we assume that the Bob and the Eve are located within the maximum transmission distance.
\subsection{Optimization Scheme}
Given the effective communication range of the system, the optimization problem in the single‑user scenario can be investigated. The optimization problem can be formulated as
\begin{subequations}\label{P2}
	\begin{align}
		\mathrm{P2}:\quad
		\max_{\mathcal{S}_1}
		\quad&
		\left[
		R_1-R_{E,1}
		\right]^+
		\label{P2a}
		\\
		\mathrm{s.t.}\quad
		&R_1\geq R_{\min}, 
		\label{P2b}\\
		&\mathbf{h}_1^H\mathbf{v}_1 = 0,
		\label{P2c}\\
		&\|\mathbf{w}_1\|^2=1,
		\label{P2d}\\
		&\|\mathbf{v}_1\|^2=1,
		\label{P2e}\\
		&0\leq\beta\leq1,
		\label{P2f}\\
		&0\leq\Delta h\leq\Delta h_{\max},
		\label{P2g}\\
		&0\leq\theta\leq\theta_{\max}, 0\leq\phi\leq\phi_{\max},
		\label{P1h}\\
		&0\leq\theta_{z,n}\leq\theta_{z,\max},
		\quad \forall n\in\mathcal{N},
		\label{P2i}
	\end{align}
\end{subequations}
where $\mathcal{S}_1
\triangleq
\left\{
\mathbf{w}_1,\mathbf{v}_1,
\beta,\Delta h,\theta,\phi,
\{\theta_{z,n},\phi_{a,n}\}_{n=1}^{N}
\right\}$. Problem P2 is highly non-convex since the precoding vectors, power allocation factor, array elevation, array rotation angles, and antenna boresight orientations are strongly coupled in both the legitimate and eavesdropping channels. To address this issue, an AO framework is adopted.
For the single-user scenario, i.e., $K=1$, we denote
$\mathbf{w}_1$ and $\mathbf{v}_1$ by $\mathbf{w}$ and $\mathbf{v}$,
respectively. It is worth noting that $N\geq 2$ is required in P2
such that the zero-forcing AN constraint
$\mathbf{h}_1^H\mathbf{v}=0$ can be satisfied with
$\|\mathbf{v}\|^2=1$.
\subsubsection{Optimization of the Information Precoding Vector}

For given $\mathbf{v}$, $\beta$, $\Delta h$, $\theta$, $\phi$,
and the antenna boresight orientations, P2 can be reformulated as
\begin{subequations}\label{P3}
	\begin{align}
		\mathrm{P3}:\quad
		\max_{\mathbf{w}}
		\quad&
		\left[
		R_1-R_{E,1}
		\right]^+
		\label{P3a}
		\\
		\mathrm{s.t.}\quad
		&R_1\geq R_{\min},
		\label{P3b}\\&\mathbf{w}^H\mathbf{w}=1.\label{P3c}
	\end{align}
\end{subequations}
P3 can be solved using semidefinite relaxation (SDR) and successive convex approximation (SCA). Specifically, by introducing an auxiliary variable $\mathbf{W}=\mathbf{w}\mathbf{w}^H$, we have
\begin{align}
	\mathbf{W}\succeq\mathbf{0},\qquad
	\operatorname{Tr}(\mathbf{W})=1,\qquad
	\operatorname{rank}(\mathbf{W})=1.
\end{align}
 Substituting $\mathbf{W}$ into the objective function, we have
\begin{align}
	R_s(\mathbf{W})
	=
	&
	\log_2 A_b(\mathbf{W})
	-
	\log_2 B_b(\mathbf{W})
	\notag\\
	&
	-
	\log_2 A_e(\mathbf{W})
	+
	\log_2 B_e(\mathbf{W}),
	\label{eq:Rs_W}
\end{align}
where
\begin{align}
&	B_b(\mathbf{W})
	=
	\beta(1+\kappa_r)\kappa_t
	\operatorname{Tr}(\mathbf{D}_b\mathbf{W})+
	\beta\kappa_r
	\operatorname{Tr}(\mathbf{h}_1\mathbf{h}_1^H\mathbf{W})
	+c_b,
	\\
&	B_e(\mathbf{W})
	=
	\beta\kappa_t
	\operatorname{Tr}(\mathbf{D}_e\mathbf{W})
	+c_e,
	\\
&	A_b(\mathbf{W})
	=
	B_b(\mathbf{W})
	+
	\beta\operatorname{Tr}(\mathbf{h}_1\mathbf{h}_1^H\mathbf{W}),\\
&	A_e(\mathbf{W})
	=
	B_e(\mathbf{W})
	+
	\beta\operatorname{Tr}(\mathbf{h}_e\mathbf{h}_e^H\mathbf{W}),\\
	&	\mathbf{D}_b
	=\operatorname{diag}
	\left(|\mathbf{h}_1[1]|^2,\ldots,|\mathbf{h}_1[N]|^2\right),\\
&\mathbf{D}_e
	=\operatorname{diag}
	\left(|\mathbf{h}_e[1]|^2,\ldots,|\mathbf{h}_e[N]|^2\right),\\
&	c_b
	=
	(1+\kappa_r)\kappa_t(1-\beta)
	\mathbf{v}^H\mathbf{D}_b\mathbf{v}
	+
	(1+\kappa_r)\frac{\sigma^2}{P_0},\\
&	c_e
	=
	(1-\beta)\mathbf{v}^H\mathbf{h}_e\mathbf{h}_e^H\mathbf{v}
	+
	\kappa_t(1-\beta)
	\mathbf{v}^H\mathbf{D}_e\mathbf{v}
	+
	\frac{\sigma^2}{P_0}.
\end{align}
Note that the zero-forcing AN constraint satisfies
$\mathbf{h}_1^H\mathbf{v}=0$. Moreover, the hardware-impairment-related term can be equivalently expressed as
\begin{align}
	\mathbf{h}_1^H
	\left(
	(\mathbf{w}\mathbf{w}^H)\odot\mathbf{I}_N
	\right)
	\mathbf{h}_1
	=
	\mathbf{w}^H
	\operatorname{diag}
	\left(
	|h_1[1]|^2,\ldots,|h_1[N]|^2
	\right)
	\mathbf{w}.
\end{align}

Since \eqref{eq:Rs_W} is non-concave, the SCA method is employed. Since $\ln(x)$ is concave,
for any $x>0$, its first-order Taylor expansion provides an upper
bound as
$
	\ln x
	\leq
	\ln x^{(r)}
	+
	\frac{x-x^{(r)}}{x^{(r)}},
$
where $x^{(r)}$ denotes the value obtained at the $r$-th iteration.
At the $r$-th iteration, the following lower bounds hold:
\begin{align}
	-\ln B_b(\mathbf{W})
	\geq
	&
	-\ln B_b^{(r)}
	-
	\frac{
		B_b(\mathbf{W})-B_b^{(r)}
	}{
		B_b^{(r)}
	},
	\\
	-\ln A_e(\mathbf{W})
	\geq
	&
	-\ln A_e^{(r)}
	-
	\frac{
		A_e(\mathbf{W})-A_e^{(r)}
	}{
		A_e^{(r)}
	},
\end{align}
where $B_b^{(r)}=B_b(\mathbf{W}^{(r)})$ and
	$A_e^{(r)}=A_e(\mathbf{W}^{(r)})$.
Accordingly, a concave lower bound of the SR can be
constructed as
\begin{align}
	\widetilde{R}_s(\mathbf{W})
	=
	\frac{1}{\ln 2}
	\Bigg[
	&
	\ln A_b(\mathbf{W})
	+
	\ln B_e(\mathbf{W})
	\notag\\
	&
	-\ln B_b^{(r)}
	-
	\frac{
		B_b(\mathbf{W})-B_b^{(r)}
	}{
		B_b^{(r)}
	}
	\notag\\
	&
	-\ln A_e^{(r)}
	-
	\frac{
		A_e(\mathbf{W})-A_e^{(r)}
	}{
		A_e^{(r)}
	}
	\Bigg].
\end{align}
Then, the QoS constraint $R_1\geq R_{\min}$ is equivalent to
\begin{align}
	\beta\operatorname{Tr}(\mathbf{h}_1\mathbf{h}_1^H\mathbf{W})
	\geq
	(2^{R_{\min}}-1) B_b(\mathbf{W}),
\end{align}
which is an affine constraint with respect to $\mathbf{W}$.
Therefore, the information precoding optimization problem can be transformed into
\begin{subequations}
	\begin{align}
		\mathrm{P4}:\quad
		\max_{\mathbf{W}}\quad&
		\widetilde{R}_s(\mathbf{W})
		\\
		\mathrm{s.t.}\quad
		&
		\beta\operatorname{Tr}(\mathbf{h}_1\mathbf{h}_1^H\mathbf{W})
		\geq
		(2^{R_{\min}}-1)B_b(\mathbf{W}),
		\\
		&
		\operatorname{Tr}(\mathbf{W})=1,
		\\
		&
		\mathbf{W}\succeq\mathbf{0}.
	\end{align}
\end{subequations}
The rank-one constraint is relaxed, and thus P3 becomes a convex semidefinite programming (SDP) problem, which can be solved by standard convex optimization tools. If the obtained $\mathbf{W}^{\star}$ is rank one, $\mathbf{w}^{\star}$ can be directly recovered by eigenvalue decomposition. Otherwise, Gaussian randomization can be employed to obtain a feasible rank-one solution.

\subsubsection{Optimization of the AN Precoding Vector}

For fixed $\mathbf{W}$, $\beta$, $\Delta h$, $\theta$, $\phi$,
and the antenna boresight orientations, 
the optimization subproblem with respect to $\mathbf{v}$ can be formulated as
\begin{subequations}\label{P5}
	\begin{align}
		\mathrm{P5}:\quad
		\max_{\mathbf{v}}
		\quad&
-R_{E,1}
		\label{P5a}
		\\
		\mathrm{s.t.}\quad
		&\mathbf{h}_1^H\mathbf{v} = 0,
\label{P5b}\\&\mathbf{v}^H\mathbf{v}=1.\label{P5c}
	\end{align}
\end{subequations}
Ignoring the constant terms, the objective function \eqref{P5a} in P5 can be rewritten as A. For constraint \eqref{P5b}, zero‑forcing AN beamforming can be exploited. Specifically, a projection matrix is introduced, denoted as 
$
	\mathbf{P}_b^{\perp}
	=
	\mathbf{I}_N
	-
	\frac{
		\mathbf{h}_1\mathbf{h}_1^H
	}{
		\|\mathbf{h}_1\|^2
	}.
$
By introducing an auxiliary variable $\bm{\varrho}$, the AN precoding vector can be expressed as $\mathbf{v} = \mathbf{P}_b^{\perp}\bm{\varrho}$. Therefore, we have
\begin{subequations}\label{P6}
	\begin{align}
		\mathrm{P6}:
		\max_{\bm{\varrho}}
		\ &
\bm{\varrho}^H{(\mathbf{P}_b^{\perp})}^H(\mathbf{h}_e\mathbf{h}_e^H+\kappa_t\mathbf{D}_e-\nu_{\mathrm{aux}}\mathbf{D}_b)\mathbf{P}_b^{\perp}\bm{\varrho}
		\label{P6a}
		\\
		\mathrm{s.t.}\ &\bm{\varrho}^H{(\mathbf{P}_b^{\perp})}^H\mathbf{P}_b^{\perp}\bm{\varrho}=1,\label{P6b}
	\end{align}
\end{subequations}
$\nu_{\mathrm{aux}}$ denotes the introduced auxiliary variable. To further account for the AN-induced transmitter distortion at Bob, we construct an auxiliary AN design criterion by introducing a nonnegative weighting factor $\nu_{\mathrm{aux}}$. Since $\mathbf{P}_b^{\perp}$ is a singular matrix, by performing singular value decomposition (SVD) $\mathbf{P}_b^{\perp}=\mathbf{U}_r\mathbf{S}\mathbf{V}_r^H$, we obtain $\bm{\varrho}=\mathbf{U}_r\tilde{\bm{\varrho}}$, $\mathbf{A}=\mathbf{U}_r^H{(\mathbf{P}_b^{\perp})}^H(\mathbf{h}_e\mathbf{h}_e^H+\kappa_t\mathbf{D}_e-\nu_{\mathrm{aux}}\mathbf{D}_b)\mathbf{P}_b^{\perp}\mathbf{U}_r$, and $\mathbf{B}=\mathbf{U}_r^H{(\mathbf{P}_b^{\perp})}^H\mathbf{P}_b^{\perp}\mathbf{U}_r$. Then, we have
\begin{subequations}\label{P7}
	\begin{align}
		\mathrm{P7}:
		\max_{\tilde{\bm{\varrho}}}
		\ &
	\frac{\tilde{\bm{\varrho}}^H\mathbf{A}\tilde{\bm{\varrho}}}{\tilde{\bm{\varrho}}^H\mathbf{B}\tilde{\bm{\varrho}}}
		\label{P7a}
		\\
		\mathrm{s.t.}\ &\tilde{\bm{\varrho}}^H\tilde{\bm{\varrho}}=1,\label{P7b}
	\end{align}
\end{subequations}
P7 is a generalized Rayleigh quotient problem, and its solution is the eigenvector corresponding to the largest eigenvalue of $\mathbf{B}^{-1}\mathbf{A}$.
\subsubsection{Optimization of the Power Allocation Factor}

For fixed $\mathbf{W}$, $\mathbf{v}$, $\Delta h$, $\theta$, $\phi$, and the antenna boresight orientations, the optimization subproblem with respect to $\beta$ can be formulated as
\begin{subequations}\label{P8}
	\begin{align}
		\mathrm{P8}:
		\max_{\beta}
		\ &
		\log_2(1+\frac{\beta a_1}{a_2+\beta a_3})-\log_2(1+\frac{\beta b_1}{b_2+\beta b_3})
		\label{P8a}
		\\
		\mathrm{s.t.}\ &\log_2(1+\frac{\beta a_1}{a_2+\beta a_3})\geq R_{\min},\label{P8b}\\
				&0\leq\beta\leq1,
		\label{P8c}
	\end{align}
\end{subequations}
where
\begin{align}
	a_1
	&=
	\operatorname{Tr}(\mathbf{h}_1\mathbf{h}_1^H\mathbf{W}),
	\\
	a_2
	&=
	(1+\kappa_r)\kappa_t
	\mathbf{v}^H\mathbf{D}_b\mathbf{v}
	+
	(1+\kappa_r)\frac{\sigma^2}{P_0},
	\\
	a_3
	&=
	(1+\kappa_r)\kappa_t
	\left[
	\operatorname{Tr}(\mathbf{D}_b\mathbf{W})
	-
	\mathbf{v}^H\mathbf{D}_b\mathbf{v}
	\right]+
	\kappa_r
	\operatorname{Tr}(\mathbf{h}_1\mathbf{h}_1^H\mathbf{W}),
\\
	b_1
	&=
	\operatorname{Tr}(\mathbf{h}_e\mathbf{h}_e^H\mathbf{W}),
\end{align}\begin{align}
	b_2
	&=
	\mathbf{v}^H\mathbf{h}_e\mathbf{h}_e^H\mathbf{v}
	+
	\kappa_t\mathbf{v}^H\mathbf{D}_e\mathbf{v}
	+
	\frac{\sigma^2}{P_0},
	\\
	b_3
	&=
	\kappa_t\operatorname{Tr}(\mathbf{D}_e\mathbf{W})
	-
	\mathbf{v}^H\mathbf{h}_e\mathbf{h}_e^H\mathbf{v}
	-
	\kappa_t\mathbf{v}^H\mathbf{D}_e\mathbf{v}.
\end{align}
We first determine the feasible interval of $\beta$ imposed by the QoS constraint. 
Then, constraint \eqref{P8b} is equivalent to
$	\frac{\beta a_1}{a_2+\beta a_3}\geq2^{R_{\min}}-1.$
 Since $a_2+\beta a_3>0$, the above inequality can be rewritten as
$
	\beta\geq\frac{(2^{R_{\min}}-1)a_2}{\left(a_1-(2^{R_{\min}}-1)a_3\right)}$.
Hence, a feasible solution exists only if $a_1-(2^{R_{\min}}-1)a_3>0$.
Under this condition, the minimum feasible power allocation factor is given by
$\beta_{\min}=\frac{(2^{R_{\min}}-1)a_2}{a_1-(2^{R_{\min}}-1)a_3}$.
Therefore, provided that $\beta_{\min}\leq1$, the feasible interval of the power allocation factor is
$
	\beta\in[\beta_{\min},1].$
If either $a_1-(2^{R_{\min}}-1)a_3\leq0$ or $\beta_{\min}>1$, P8 is infeasible under the given $\mathbf{W}$, $\mathbf{v}$, and antenna configuration. Next, the objective function in \eqref{P8a} can be equivalently rewritten as
\begin{align}
	f(\beta)=\log_2\left(\frac{a_2+\beta(a_1+a_3)}{a_2+\beta a_3}\right)-\log_2\left(\frac{b_2+\beta(b_1+b_3)}{b_2+\beta b_3}\right).
\end{align}
Equivalently, we have
\begin{align}
	f(\beta)=&\log_2\left[a_2+\beta(a_1+a_3)\right]-\log_2(a_2+\beta a_3)\notag\\&-\log_2\left[b_2+\beta(b_1+b_3)\right]+\log_2(b_2+\beta b_3).
\end{align}
Taking the first-order derivative of $f(\beta)$ with respect to $\beta$ yields
\begin{align}\label{eq:df_beta}
	f'(\beta)=&\frac{1}{\ln 2}\big[\frac{a_1a_2}{\left[a_2+\beta(a_1+a_3)\right](a_2+\beta a_3)}\notag\\&-\frac{b_1b_2}{\left[b_2+\beta(b_1+b_3)\right](b_2+\beta b_3)}\big].
\end{align}
The stationary points in the interior of the feasible interval can therefore be obtained by setting $f'(\beta)=0$.
After expanding and collecting the terms associated with $\beta$, the above equation can be expressed as
\begin{equation}\label{eq:beta_quad}
	c_2\beta^2+c_1\beta+c_0=0,
\end{equation}
where
	$c_0=a_2b_2(a_1b_2-b_1a_2)$,
	$c_1=2a_2b_2(a_1b_3-b_1a_3)$,
and	$c_2=a_1a_2b_3(b_1+b_3)-b_1b_2a_3(a_1+a_3)$.
When $c_2\neq0$ and $c_1^2-4c_2c_0\geq0$, the two stationary-point candidates are given by
\begin{equation}\label{eq:beta_roots}
	\beta_{1,2}=\frac{-c_1\pm\sqrt{c_1^2-4c_2c_0}}{2c_2}.
\end{equation}
Only the real-valued roots satisfying $\beta_i\in[\beta_{\min},1]$ are retained as feasible stationary points. When $c_2=0$ and $c_1\neq0$, \eqref{eq:beta_quad} reduces to a linear equation and the stationary point is given by
$
	\beta_0=-\frac{c_0}{c_1},
$
which is retained only if $\beta_0\in[\beta_{\min},1]$. If $c_2=c_1=0$ but $c_0\neq0$, no interior stationary point exists, whereas if $c_2=c_1=c_0=0$, the derivative vanishes over the feasible interval and any feasible $\beta$ achieves the same objective value. Since $f(\beta)$ is continuous over the closed feasible interval $[\beta_{\min},1]$, its global optimum must occur either at a feasible stationary point or at one of the interval boundaries. Accordingly, define the candidate set as
\begin{equation}
	\mathcal{B}=\left\{\beta_{\min},1,\beta_i\mid f'(\beta_i)=0,\ \beta_i\in[\beta_{\min},1]\right\}.
\end{equation}
The globally optimal power allocation factor for P8 can then be obtained as
\begin{equation}\label{eq:beta_opt}
	\beta^\star=\operatorname*{arg\,max}_{\beta\in\mathcal{B}}f(\beta).
\end{equation}
Therefore, P8 can be solved in closed form up to the comparison of a finite number of candidate points, and at most two feasible interior stationary points together with the two boundary points need to be evaluated.
\subsubsection{Optimization of the Antenna Configuration}

Define the antenna configuration as
\begin{align}
	\mathcal{Q}
	=
	\{
	\Delta h,\theta,\phi,
	\theta_{z,1},\ldots,\theta_{z,N},
	\phi_{a,1},\ldots,\phi_{a,N}
	\}.
\end{align}
The antenna configuration variables affect both the legitimate and eavesdropping channels through the path gain, propagation direction, array response, and directional antenna gain. 
For fixed $\mathbf{W}$, $\mathbf{v}$, and $\beta$, the antenna
configuration optimization problem is formulated as
\begin{subequations}\label{P9}
	\begin{align}
		\mathrm{P9}:\quad
		\max_{\mathcal{Q}}\quad&
		\left[
		R_1(\mathcal{Q})-R_{E,1}(\mathcal{Q})
		\right]^+
		\\
		\mathrm{s.t.}\quad
		&
		R_1(\mathcal{Q})\geq R_{\min},
		\\
		&
		0\leq\Delta h\leq\Delta h_{\max},
		\\
		&
		0\leq\theta\leq\theta_{\max},	0\leq\phi\leq\phi_{\max},
		\\
		&0\leq\theta_{z,n}\leq\theta_{z,\max},
		\quad \forall n.
	\end{align}
\end{subequations}
Due to the highly nonlinear coupling among the antenna configuration variables, P9 is difficult to solve in closed form. For physical-layer security, the distinguishability between the legitimate and eavesdropping channels plays an important role in determining the secrecy performance. A lower spatial correlation between the Bob's channel and the Eve's channel generally provides better spatial separability, thereby facilitating the suppression of information leakage and potentially improving the achievable SR. Based on this idea, we performed antenna configuration optimization.

First, we consider orienting the array toward the Bob. Based on \eqref{34} and \eqref{39}, when serving long‑distance Bob, we consider increasing the array height $\Delta h$ while satisfying the QoS requirement. When serving short‑distance Bob, the array height is kept unchanged. This is due to the consideration that user QoS is of high priority. Then, antenna orientation optimization is considered. The antenna orientations are initially directed toward Bob to provide the potential maximum gain. Based on the alternating optimization principle, the antenna orientations are aligned with the optimized channel gains corresponding to Bob. The optimized channel gain corresponding to Bob is defined as \begin{align}
	\bm{\varpi}=|\mathbf{h}_1^*\odot\mathbf{w}|.
\end{align}
To preserve the relative gain profile while fully exploiting the available antenna gain, the desired directional gain of the $n$-th antenna toward Bob is defined as \begin{align}\bar G_{n,b}=\frac{G_0(\bm{\varpi}[n])^2}{\|\bm{\varpi}\|^2_{\infty}}.\end{align} Optimizing the directivity of the antenna radiation pattern may affect the achievable transmission rate at Eve. To address this issue, we first derive the adjustable range $[{G}_{n,\min},{G}_{n,\max}]$ of the directional gain of the $n$-th antenna at Eve, and then minimize the achievable transmission rate at Eve within this adjustable range. Define $\mathcal{G} = \{G_{1,e},G_{2,e},\ldots,G_{N,e}\}$. The corresponding optimization problem can be formulated as 
\begin{subequations}\label{P10}
	\begin{align}
		\mathrm{P10}:\quad
		\max_{\mathcal{G}}\quad&
		-R_{E,1}(\mathcal{G})
		\\
		\mathrm{s.t.}\quad
		&
		{G}_{n,\min}\leq G_{n,e}\leq{G}_{n,\max},\forall n\in\mathcal{N}.
	\end{align}
\end{subequations}
where
\begin{align}
	&G_{n,\min}=G_0\left[\max\left\{0,\cos(\alpha_n+\delta_n)\right\}\right]^{2p},\\
	&G_{n,\max}=G_0\left[\max\left\{0,\cos|\alpha_n-\delta_n|\right\}\right]^{2p},\\
	&\alpha_n=\arccos\left[\left(\frac{\bar G_{n,b}}{G_0}\right)^{\frac{1}{2p}}\right],\end{align}\begin{align}
	&\delta_n=\arccos\left(\mathbf{d}_{n,b}^{T}\mathbf{d}_{n,e}\right),\\
	&\mathbf{d}_{n,b}=\frac{\mathbf{b}_1-\mathbf{p}_n^{\prime\prime}}{\|\mathbf{b}_1-\mathbf{p}_n^{\prime\prime}\|},
	\qquad
	\mathbf{d}_{n,e}=\frac{\mathbf{e}-\mathbf{p}_n^{\prime\prime}}{\|\mathbf{e}-\mathbf{p}_n^{\prime\prime}\|}.
\end{align}
In particular, when $\alpha_n+\delta_n\geq\pi/2$, the boresight can be adjusted such that Eve lies outside the effective radiation region, yielding $G_{n,\min}=0$. P10 can be rewritten as
\begin{subequations}\label{P11}
	\begin{align}
		\mathrm{P11}:\ 
		\max_{\mathbf{g}}\ &
	\frac{\mathbf{g}^T\mathbf{B}_e\mathbf{g}}{\mathbf{g}^T(\mathbf{A}_e+\mathbf{B}_e)\mathbf{g}}\\
		\mathrm{s.t.}\ 
		&
		\sqrt{{G}_{n,\min}}\leq \mathbf{g}[n]\leq\sqrt{{G}_{n,\max}},\forall n\in\mathcal{N},
	\end{align}
\end{subequations}
where \begin{align}
	&\mathbf{g}=[\sqrt{G_{1,e}},\sqrt{G_{2,e}},\ldots,\sqrt{G_{N,e}}]^T,\\
	&\mathbf{A}_e = \beta(\tilde{\mathbf{h}}_e^*\odot\mathbf{w})
	(\tilde{\mathbf{h}}_e^*\odot\mathbf{w})^H,\\&\mathbf{B}_e =	(1-\beta)(\tilde{\mathbf{h}}_e^*\odot\mathbf{v})
(\tilde{\mathbf{h}}_e^*\odot\mathbf{v})^H\notag\\&\qquad+(1-\beta)\kappa_t\mathrm{diag}
(|\tilde{\mathbf{h}}_e^*|^2\odot|\mathbf{v}|^2)\notag\\&\qquad+\beta\kappa_t\mathrm{diag}
(|\tilde{\mathbf{h}}_e^*|^2\odot|\mathbf{w}|^2)+\frac{\sigma^2}{P_0}\mathbf{I}_N,\\&\tilde{\mathbf{h}}_e[n] = \rho_e^{\frac{1}{2}}(\sqrt{\frac{\xi_R}{\xi_R + 1}} h_{\text{LoS}}^{n,e} + \sqrt{\frac{1}{\xi_R + 1}} h_{\text{NLoS}}^{n,e}).\end{align}
To reduce complexity, we adopt the generalized Rayleigh quotient approximation for P11. Specifically, after relaxing the constraints, the solution $\tilde{\mathbf{g}}$ for $\mathbf{g}$ is the eigenvector corresponding to the largest eigenvalue of $(\mathbf{A}_e+\mathbf{B}_e)^{-1}\mathbf{B}_e$. Then, by introducing an auxiliary variable $c_e$, we have
\begin{align}
\mathbf{g} = c_e\tilde{\mathbf{g}}.
\end{align}
According to ${\sqrt{{G}_{n,\min}}}\leq c_e\tilde{\mathbf{g}}[n]\leq{\sqrt{{G}_{n,\max}}}$, we have $c_{\min}=\displaystyle\max_{n}{\sqrt{{G}_{n,\min}}}/{\tilde{\mathbf{g}}[n]}$ and $c_{\max}=\displaystyle\min_{n}{\sqrt{{G}_{n,\max}}}/{\tilde{\mathbf{g}}[n]}$. To minimize the achievable transmission rate at Eve, when 
$c_{\min}\leq c_{\max}$, 
$c_{\min}$ is used to satisfy the constraint. When 
$c_{\min}>c_{\max}$, we have
\begin{align}
\mathbf{g}[n]
=
\min\left\{
\sqrt{G_{n,\max}},
\max\left\{
\sqrt{G_{n,\min}},
c_{\min}\tilde{\mathbf{g}}[n]
\right\}
\right\}.
\end{align}

The computational complexity of the proposed AO scheme mainly arises from the information precoder, AN precoder, power allocation, and antenna-orientation optimization. Let $I_{\rm A}$ and $I_{\rm S}$ denote the numbers of AO and SCA iterations, respectively. Solving P4 requires $\mathcal{O}(I_{\rm S}C_{\rm SDP}(N,\epsilon))$ operations, where $C_{\rm SDP}(N,\epsilon)$ denotes the complexity of solving an SDP with an $N\times N$ positive semidefinite variable. The AN precoder and antenna-orientation updates both require an eigenvalue or generalized eigenvalue decomposition with complexity $\mathcal{O}(N^3)$, whereas the power-allocation update requires $\mathcal{O}(N^2)$ operations. Therefore, the overall complexity is $\mathcal{O}\!\left[I_{\rm A}\left(I_{\rm S}C_{\rm SDP}(N,\epsilon)+N^3\right)\right]$.
\section{Multi‑User Optimization Scheme}
The preceding section investigates the single-user case and reveals the coupling among information precoding, AN precoding, power allocation, and antenna configuration. However, directly extending the AO framework to the multi-user case results in a considerably higher computational burden. In P1, $K$ information precoders and $K$ AN precoders are jointly coupled through multi-user interference and hardware impairments, while the common array elevation, array rotation, and element orientations simultaneously affect the channels of all Bobs and Eve. Consequently, repeatedly solving the corresponding optimization subproblems becomes computationally expensive as the numbers of users and antennas increase. Moreover, the optimal variables of different users are strongly correlated through the shared propagation environment, which makes independent variable optimization inefficient. To address this challenge, a Transformer‑based multi‑user scheme is proposed. For clarity, the algorithm proposed in this section is referred to as the Transformer‑based learning algorithm.

\subsection{Initialization Setup}
\label{subsec:mu_encoder}
For the initial antenna configuration $\mathcal Q^0$, we compute the channel strength and direction related to the transmit power and noise, which can be expressed as
\begin{equation}
	\bar{\mathbf h}_x^0=
	\sqrt{\frac{P_0}{K\sigma^2}}\,\mathbf h_x(\mathcal Q^0),
	\qquad
	\widehat{\mathbf h}_x^0=\mathcal U(\bar{\mathbf h}_x^0),
	\label{eq:mu_reference_channel}
\end{equation}
where $\mathcal U(\mathbf h)=\mathbf h/
\|\mathbf h\|$. The reference configuration is used only to construct input features. The channels used to evaluate each candidate are recomputed at that candidate's actual configuration. The initial parameters of the antenna height, array orientation, and antenna orientation are initialized to the midpoint of their adjustable ranges. Then,
the real-valued feature vector of receiver $x$  can be denoted as
	\begin{align}
		\mathbf f_x=\big[&\mathbf c_x^T/\ell_0,\ \iota_x/s_0,
		\ \log(1+\|\bar{\mathbf h}_x^0\|_2),\notag\\
		&\Re\{\widehat{\mathbf h}_x^0\}^T,
		\Im\{\widehat{\mathbf h}_x^0\}^T,\Re\{\boldsymbol\nu_x\}^T,
		\Im\{\boldsymbol\nu_x\}^T\big]^T,\label{eq:mu_receiver_features}
	\end{align}
	where $\mathbf c_x$ denotes the receiver's position, and $\iota_x$ is the corresponding shadowing coefficient in dB. The constants $\ell_0$ and $s_0$ normalize the position and shadowing features, respectively. The vector $\boldsymbol\nu_x$ collects the instantaneous NLoS fading coefficients associated with the $N$ antenna elements. 
Each antenna token is constructed from $\mathbf p_n/\lambda$. Separate linear embeddings map these features into a common dimension $d$, and trainable type embeddings distinguish Bobs, Eve, antennas, and the global token. Thus, the encoder input contains $L=N+K+2$ tokens.
The tokens are processed by $D_{\mathrm{enc}}$ Transformer encoder layers, each containing multi-head self-attention and a feedforward network. For one attention head, we have
\begin{equation}
	\operatorname{Att}(\mathbf Q_a,\mathbf K_a,\mathbf V_a)
	=\operatorname{softmax}\!\left(
	\frac{\mathbf Q_a\mathbf K_a^T}{\sqrt{d_a}}\right)\mathbf V_a,
	\label{eq:mu_attention}
\end{equation}
where $d_a$ is the head dimension. 

Let $\mathbf t_g$, $\mathbf t_k$, and $\mathbf t_n^{\mathrm A}$ denote the encoded global, Bob, and antenna tokens, respectively. Then, the output heads can be denoted as
\begin{equation}
		\mathbf u_g=f_g(\mathbf t_g),
		\mathbf u_n^{\mathrm A}=f_{\mathrm A}(\mathbf t_n^{\mathrm A}),
		\mathbf r_{w,k}=f_w(\mathbf t_k),
		\mathbf r_{v,k}=f_v(\mathbf t_k).
	\label{eq:mu_heads}
\end{equation}

\subsection{Physical Mapping and Constraint Handling}
\label{subsec:mu_decoder}

\subsubsection{Power Allocation and Geometry}
Let $s(u)=(1+e^{-u})^{-1}$. Then, the raw scalar outputs are mapped as
\begin{equation}
	\begin{aligned}
		&\beta=s(u_\beta), &
		&\Delta h=\Delta h_{\max}s(u_h),\\
		&\theta=\theta_{\max}s(u_\theta), &
		&\phi=\phi_{\max}s(u_\phi),\\
		&\theta_{z,n}=\theta_{z,\max}s(u_{z,n}), &
		&\phi_{a,n}=2\pi s(u_{a,n}).
	\end{aligned}
	\label{eq:mu_bounds}
\end{equation}
These mappings satisfy \eqref{P1f}--\eqref{P1i}. After every update, the channels $\mathbf h_k$ and $\mathbf h_e$ are reconstructed from the current antenna configuration.

\subsubsection{Information Precoding}
Define the normalized Bob-channel matrix and its null-space projector as
\begin{equation}
	\begin{aligned}
		\!\mathbf A=[\widehat{\mathbf h}_1,\ldots,
		\widehat{\mathbf h}_K]^H,
		\ \widehat{\mathbf h}_k=\mathcal U(\mathbf h_k),\
		\mathbf P_\perp=\mathbf I_N-\mathbf A^\dagger\mathbf A.
	\end{aligned}
	\label{eq:mu_projector}
\end{equation}
Let $\mathbf a_k^\dagger$ denote the $k$-th column of
$\mathbf A^\dagger$. The residual allows the final precoder to trade interference suppression against information leakage to Eve. Then, a normalized zero-forcing reference and the information precoder are given by
\begin{equation}
	\mathbf w_k^{\mathrm{ref}}=\mathcal U(\mathbf a_k^\dagger),
	\qquad
	\mathbf w_k=\mathcal U(\mathbf w_k^{\mathrm{ref}}+\mathbf r_{w,k}).
	\label{eq:mu_information_precoder}
\end{equation}

\subsubsection{Artificial-Noise Precoding}
Constraint \eqref{P1c} can be rewritten as
\begin{equation}
	\mathbf A\sum_{i=1}^{K}\mathbf v_i=\mathbf 0.
	\label{eq:mu_sum_constraint}
\end{equation}
An initial AN reference is chosen as
\begin{equation}
	\mathbf v^{\mathrm{ref}}=
	\mathcal U\!\left(\mathbf P_\perp\mathcal U(\mathbf h_e)\right),
	\qquad
	\mathbf v_i^{(0)}=
	\mathcal U(\mathbf v^{\mathrm{ref}}+\mathbf r_{v,i}).
	\label{eq:mu_an_initial}
\end{equation}
To handle both \eqref{eq:mu_sum_constraint} and \eqref{P1e}, perform $J_{\mathrm p}$ alternating projection steps:
	\begin{align}
		&\mathbf s^{(j)}=\sum_{i=1}^{K}\mathbf v_i^{(j)},\\
		&\widetilde{\mathbf v}_i^{(j+1)}
		=\mathbf v_i^{(j)}-
		\frac{(\mathbf I_N-\mathbf P_\perp)\mathbf s^{(j)}}{K},
		\mathbf v_i^{(j+1)}=\mathcal U(\widetilde{\mathbf v}_i^{(j+1)}).
	\end{align}

\subsection{Unsupervised Training}
\label{subsec:mu_training}


Let $\Delta R_k=R_k-R_{E,k}$. The hard minimum is approximated during training by
\begin{equation}
	r_\tau(\mathcal S)
	=-\tau\log\sum_{k=1}^{K}\exp(-\Delta R_k/\tau),\qquad \tau>0.
	\label{eq:mu_softmin}
\end{equation}
Here, $\log$ denotes the natural logarithm. It satisfies
\begin{equation}
	\min_k\Delta R_k-\tau\log K\ \leq\ r_\tau(\mathcal S)
	\ \leq\ \min_k\Delta R_k,
	\label{eq:mu_softmin_bound}
\end{equation}
and its derivative with respect to $\Delta R_k$ is proportional to $\exp(-\Delta R_k/\tau)$. Thus, users with smaller rate differences receive greater gradient weight. Unclipped rate differences are used so that negative differences can provide a learning signal when the original nonnegative SR is zero. Note that theoriginal objective \eqref{16} is still used for comparison.
The QoS and AN penalties can be denoted as
	\begin{align}
		\mathcal E_{\mathrm Q}(\mathcal S)
		&=\frac{1}{K}\sum_{k=1}^{K}[R_{\min}-R_k]_+^2,\\
		\mathcal E_{\mathrm A}(\mathcal S)
		&=\frac{1}{K}\left\|\mathbf A\sum_{i=1}^{K}\mathbf v_i\right\|_2^2.
	\end{align}
Then, the loss can be defined as 
\begin{equation}
	\mathcal L
	=-r_\tau(\mathcal D_\Xi(\mathbf u))
	+\mu_{\mathrm Q}\mathcal E_{\mathrm Q}(\mathcal D_\Xi(\mathbf u))
	+\mu_{\mathrm A}\mathcal E_{\mathrm A}(\mathcal D_\Xi(\mathbf u)).
	\label{eq:mu_loss}
\end{equation}
For a mini-batch $\mathcal B$, network training minimizes
\begin{equation}
	\widehat{\mathcal J}_{\mathcal B}(\boldsymbol\psi)
	=\frac{1}{|\mathcal B|}\sum_{\Xi\in\mathcal B}
	\mathcal L\big(F_{\boldsymbol\psi}(\Xi);\Xi\big).
	\label{eq:mu_batch_loss}
\end{equation}
Here, the nonnegative coefficients $\mu_{\mathrm Q}$ and $\mu_{\mathrm A}$ weight the QoS and AN constraint penalties, respectively. Moreover, $\mathbf u$ collects the raw network outputs, $\mathcal D_\Xi$ is the physical decoding map, and $F_{\boldsymbol\psi}$ denotes the Transformer mapping parameterized by the trainable weights $\boldsymbol\psi$. $\mathcal B$ denotes a mini-batch of network realizations, $|\mathcal B|$ denotes its cardinality, and $\widehat{\mathcal J}_{\mathcal B}$ denotes the corresponding empirical average loss.

\subsection{Online Joint Refinement}
\label{subsec:mu_refinement}
Adam updates optimization variables according to
\begin{equation}
	\mathbf u^{(t+1)}=\mathbf u^{(t)}
	-\alpha_{\mathrm r}
	\frac{\widehat{\mathbf m}^{(t+1)}}
	{\sqrt{\widehat{\mathbf q}^{(t+1)}}+\epsilon_{\mathrm{opt}}},
	\label{eq:mu_adam_refine}
\end{equation}
where division is elementwise and $\widehat{\mathbf m}$ and
$\widehat{\mathbf q}$ are the bias-corrected first- and second-moment estimates of the clipped loss gradients.  Every iteration updates the raw entries jointly and recomputes the physical channels and projected precoders. 
Define the principal residuals as
	\begin{align}	\label{eq:mu_residuals}
		e_{\mathrm Q}&=\max_k[R_{\min}-R_k]_+,\\
		e_{\mathrm A}&=\left\|\mathbf A\sum_i\mathbf v_i\right\|_\infty,\\
		e_{\mathrm N}&=\max_k\left\{
		\big|\|\mathbf w_k\|_2-1\big|,
		\big|\|\mathbf v_k\|_2-1\big|\right\}.
	\end{align}
Let $e_{\mathrm B}$ denote the largest nonnegative violation of the scalar bounds in \eqref{eq:mu_bounds}, with angles measured in radians, and let $\chi_{\mathrm{dom}}$ indicate that all receivers lie within the assumed channel-model and visibility ranges. A candidate is accepted as numerically feasible only if
\begin{equation}
	\begin{aligned}
		e_{\mathrm Q}\leq\epsilon_{\mathrm Q},\ 
		e_{\mathrm A}\leq\epsilon_{\mathrm A}, \
		e_{\mathrm N}\leq\epsilon_{\mathrm N},\
		e_{\mathrm B}\leq\epsilon_{\mathrm B},\ 
		\chi_{\mathrm{dom}}=1.
	\end{aligned}
	\label{eq:mu_feasibility}
\end{equation}
In the individual-nulling variant, $e_{\mathrm A}$ is replaced by $\max_{k,i}|\widehat{\mathbf h}_k^H\mathbf v_i|$. The finite QoS penalty and finite projection budget do not themselves guarantee these conditions.

An incumbent is retained separately for each scene. Feasible candidates are preferred to infeasible ones, and two feasible candidates are compared using the exact SR in (16). If neither is feasible, the implementation prefers the smaller diagnostic score
\begin{equation}
	\mathcal V=e_{\mathrm Q}+e_{\mathrm A}+e_{\mathrm N}+e_{\mathrm B}
	+(1-\chi_{\mathrm{dom}}).
	\label{eq:mu_violation_score}
\end{equation}


\subsection{Computational Complexity}
\label{subsec:mu_complexity}

For $L=N+K+2$ tokens and feed forward width proportional to $d$, the encoder, feature embeddings, and output heads have per-scene cost
$
	C_{\mathrm T}=\mathcal O\!\left(
	D_{\mathrm{enc}}(Ld^2+L^2d)+KNd\right).
	\label{eq:mu_transformer_cost}
$
The implementation explicitly constructs the dense projector $\mathbf P_\perp$. For $K<N$, the pseudoinverse costs $\mathcal O(NK^2)$, while forming the projector and applying it to $K$ reference directions costs $\mathcal O(N^2K)$. Each summed-AN projection step costs $\mathcal O(N^2+NK)$, and evaluating all receiver--precoder products costs $\mathcal O(NK^2)$. Thus, a physical mapping and rate evaluation has cost
$
	C_{\mathrm P}=\mathcal O\!\left(
	N^2K+NK^2+J_{\mathrm p}(N^2+NK)\right).
	\label{eq:mu_physical_cost}
$
With a fixed numerical-rank regime, differentiation through these operations has the same leading orders. The total online cost is
$
	C_{\mathrm{online}}=
	\mathcal O\!\left(C_{\mathrm T}+(I_{\max}+1)C_{\mathrm P}\right),
	\label{eq:mu_online_cost}
$
up to the constant number of candidate evaluations per update. For $T_{\mathrm{tr}}$ mini-batch updates of size $B_{\mathrm{tr}}$, the leading offline training cost is $\mathcal O(T_{\mathrm{tr}}B_{\mathrm{tr}}(C_{\mathrm T}+C_{\mathrm P}))$. 
\section{Simulation Results}
In this section, simulation results are presented. Unless otherwise stated, the simulation parameters are kept consistent. The specific parameter settings are as follows~\cite{Maliatsos2006}. The carrier frequency is 3 GHz. The number of antennas is $N=8$. The initial array height is $H_0 = 20$ m, and the maximum allowable increase is $\Delta h = 30$ m. The receiver height is 5 m. The maximum elevation and azimuth angles of the array are both $90^{\circ}$. For each antenna element, the maximum adjustable zenith angle is $\theta_{z,\max}=60^{\circ}$. The antenna directivity factor is $p=2$. The reference distance is 1 m. The path loss exponent is 2.2. The shadow fading is 2 dB. The noise power is -95 dBm. The same hardware impairment coefficient is assumed for the transceivers, i.e., $\kappa_t = \kappa_r = 0.01$. The minimum rate for Bobs is 1 bps/Hz. The distances from the Bobs and the Eve to the BS are randomly selected within the ranges $[800, 1600]$ and $[700, 1800]$, respectively, in meters. In addition, the main parameters of the Transformer‑based optimization scheme are listed in Table \ref{tab:transformer_settings}.
\begin{table}[t]
	\centering
	\caption{Main Parameters in the Transformer‑Based Optimization Scheme}
	\label{tab:transformer_settings}
	\footnotesize
	\renewcommand{\arraystretch}{1.15}
	\setlength{\tabcolsep}{4pt}
	\begin{tabular}{p{0.53\columnwidth}p{0.39\columnwidth}}
		\hline
		Parameter & Value \\
		\hline
		Embedding dimension, $d$
		& $64$ \\
		Number of encoder layers
		& $2$ \\
		Number of attention heads
		& $4$ \\
		Offline training updates, $T_{\mathrm{tr}}$
		& $800$ \\
		Mini-batch size
		& $32$ \\
		Offline learning rate
		& $3\times10^{-4}$ \\
		Soft-min temperature, $\tau$
		& $0.15$ \\
		AN projection iterations, $J_{\mathrm p}$
		& $40$ \\
		Online optimization iterations, $I_{\max}$
		& $200$ \\
		Online learning rate
		& $0.025$ \\
		Optimizer numerical constant, $\epsilon_{\mathrm{opt}}$
		& $10^{-8}$ \\
		\hline
	\end{tabular}
\end{table}

\begin{figure}[!t]
	\centering
	\subfloat[]{%
		\includegraphics[width=0.22\textwidth,
		trim=10 0 10 10,clip]{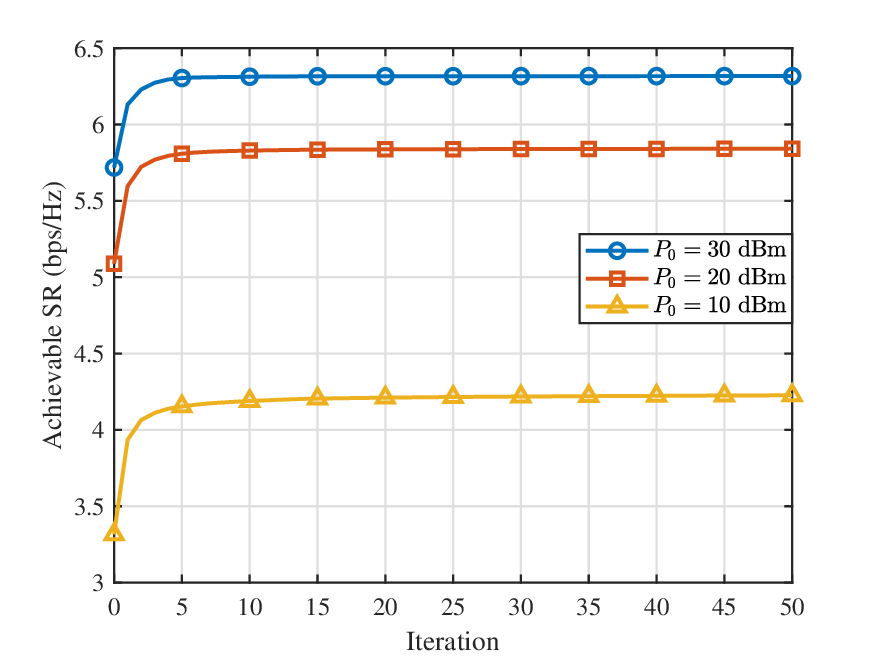}
		\label{fig:3-1}
	}
	\hfill
	\subfloat[]{%
		\includegraphics[width=0.22\textwidth,
		trim=10 0 10 10,clip]{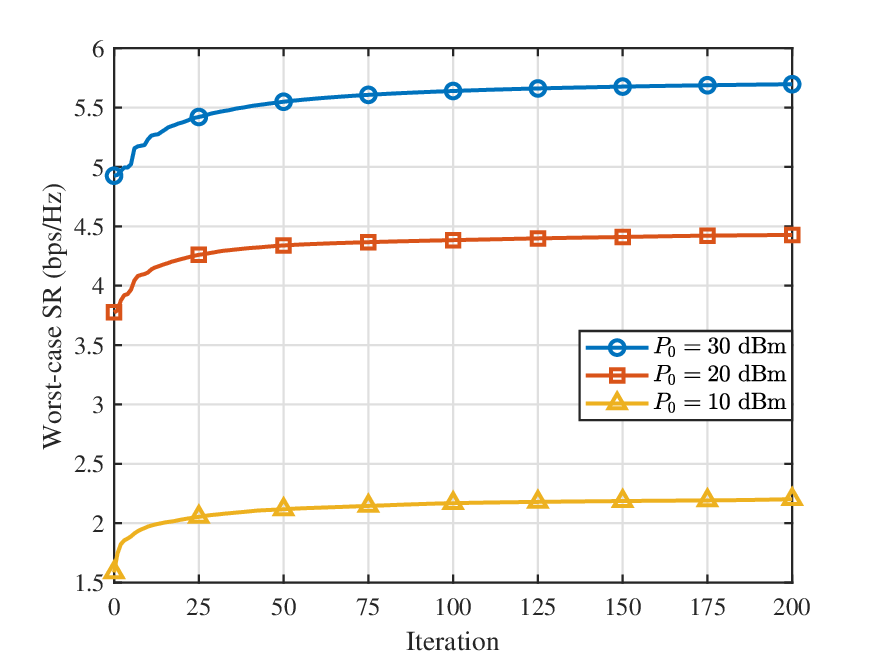}
		\label{fig:3-2}
	}
	\caption{Illustration of the iteration curves:
		(a) single‑user optimization scheme;
		(b) multi‑user optimization scheme.}
	\label{fig:3}
\end{figure}
Fig. \ref{fig:3} presents the convergence curves of the single‑user and multi‑user schemes under different transmit powers, as illustrated in \ref{fig:3-1} and \ref{fig:3-2}, respectively. Compared with the single‑user scheme, the multi‑user scheme requires more iterations to reach a stable state, since the multi‑user scenario not only requires simultaneous adjustment of multiple information precoding and AN precoding, but also needs to account for inter‑user interference, worst‑user performance, and the common antenna configuration. 

\begin{figure}[t]
	\centering
	\includegraphics[width=0.47\textwidth, trim = 0 0 0 10,clip]{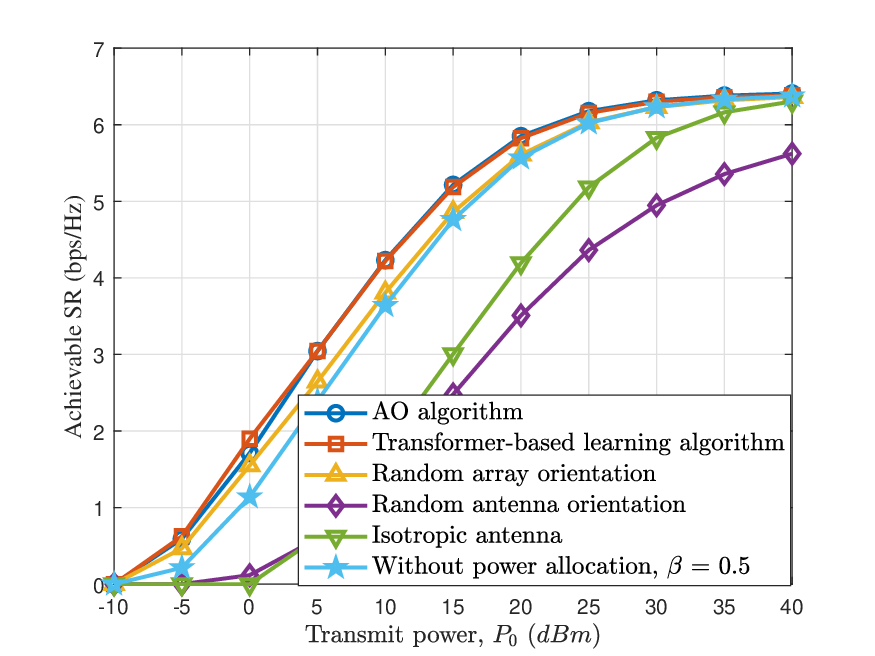}\\
	\caption{Illustration of the achievable SR versus transmit power.}\label{fig:4}
\end{figure}
Fig. \ref{fig:4} illustrates the SR versus the transmit power. As the transmit power increases, the SR value gradually improves until it reaches saturation. At low power, the system is mainly limited by thermal noise, so increasing $P_0$ significantly improves Bob’s effective SINR. As $P_0$ continues to increase, the hardware distortion power at both the transmitter and receiver grows accordingly, so the gain from further increasing the power gradually diminishes. The AO algorithm and the Transformer‑based learning algorithm achieve higher SR values and clearly outperform the other baselines. This indicates that jointly optimizing beamforming, power allocation, and the antenna physical configuration can effectively enhance the channel disparity between Bob and Eve. The random antenna orientation scheme performs poorly, because unoptimized antenna pointing may simultaneously reduce Bob’s effective array gain. Isotropic antennas are also limited in security performance since they cannot exploit directional gain to provide additional spatial discrimination between Bob and Eve. In addition, the scheme with fixed $\beta$ is overall inferior to the joint power optimization scheme, indicating that dynamically adjusting the power ratio between the information signal and AN according to channel conditions plays a role in improving the SR.

\begin{figure}
	\centering
	\includegraphics[width=0.47\textwidth, trim = 0 0 0 10,clip]{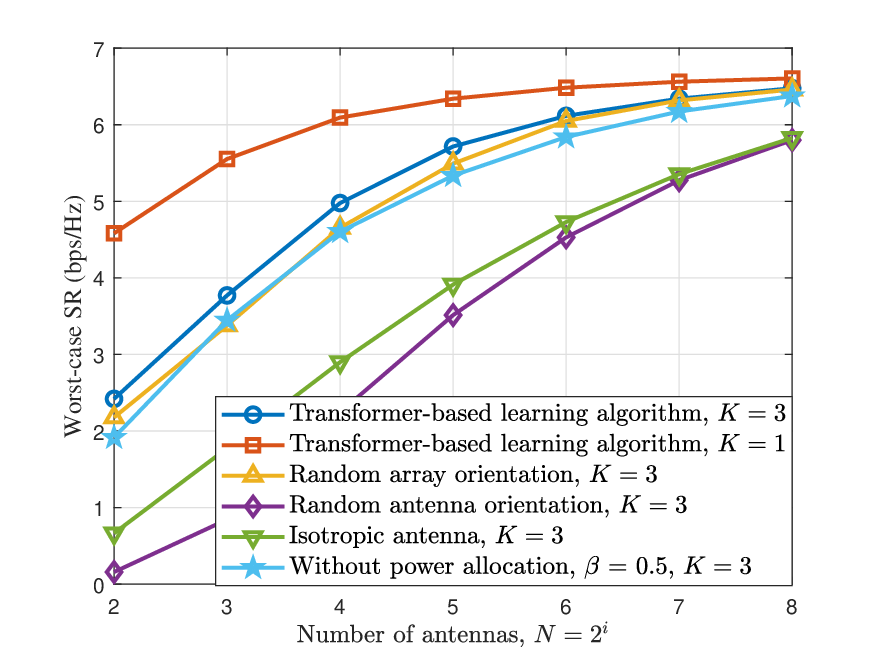}\\
	\caption{Illustration of the worst-case SR versus number of antennas.}\label{fig:5}
\end{figure}

Fig. \ref{fig:5} illustrates the trend of the worst‑case SR as the number of antennas increases. It can be seen that the worst‑case SR of both the proposed scheme and the baselines improves significantly with increasing $N$. Here, $i$ denotes the index on the horizontal axis. As $N$ increases, the system possesses more spatial degrees of freedom, and its capability to suppress multi‑user interference is enhanced, so the performance gap between $K=1$ and $K=3$ gradually narrows.
For the schemes with $K=3$, the Transformer‑based learning algorithm generally achieves a higher worst‑case user SR, while the random array orientation, random antenna orientation, and isotropic antenna schemes exhibit relatively lower performance. This indicates that, under multi‑user conditions, jointly modeling the coupling relationships among the channel, user locations, and antenna states using the Transformer‑based learning scheme is effective. In particular, when the number of antennas is small, the proposed scheme can exploit the limited spatial degrees of freedom to enhance system security performance.

\begin{figure}
	\centering
	\includegraphics[width=0.47\textwidth, trim = 0 0 0 10,clip]{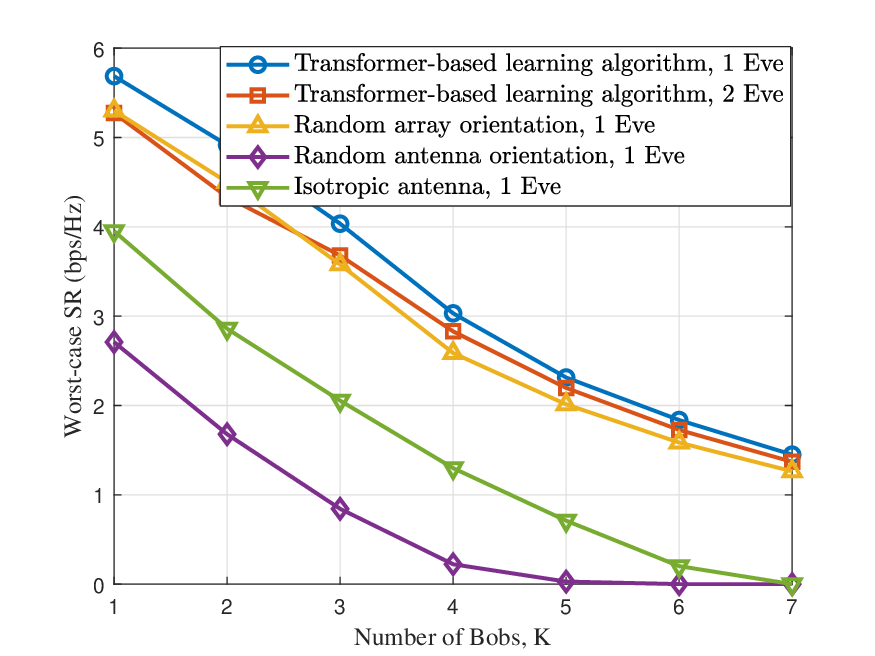}\\
	\caption{Illustration of the worst-case SR versus number of antennas.}\label{fig:6}
\end{figure}

The impact of the number of legitimate users on the worst‑case user SR is shown in Fig. \ref{fig:6}. As the number of Bobs increases, the worst‑case SR of all schemes decreases. First, under a limited transmit power, the power allocated to each user decreases as the number of users grows. Second, more users introduce more severe multi‑user interference.
The Transformer‑based learning algorithm maintains relatively good performance across different numbers of users, because the self‑attention mechanism can adjust the information beams, AN, and antenna configuration according to the channel relationships among different Bobs. In contrast, the random antenna orientation scheme degrades most rapidly as the number of users increases, and its SR approaches zero when the number of users is large. This is because random orientations cannot simultaneously provide effective directional gain for multiple Bobs and cannot effectively suppress Eve. The isotropic scheme also decreases markedly with increasing $K$, indicating that relying solely on digital beamforming is insufficient to fully exploit the additional spatial degrees of freedom offered by rotatable antennas.
In addition, the cases with one Eve and two Eves are compared. Compared with a single Eve, the worst‑case SR of the Transformer scheme is further reduced. This is because, under limited system resources, more potential eavesdropping links must be suppressed simultaneously, making it more difficult to satisfy the QoS requirements.
\begin{figure}
	\centering
	\includegraphics[width=0.47\textwidth, trim = 0 0 0 10,clip]{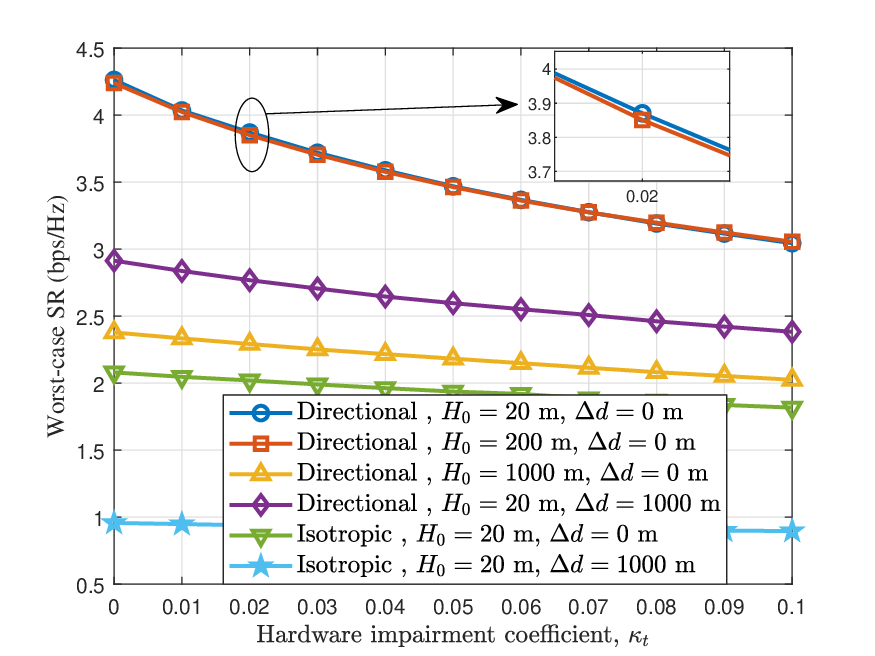}\\
	\caption{Illustration of the worst-case SR versus hardware impairment coefficient.}\label{fig:7}
\end{figure}

The variation of the worst‑case SR with the transmitter hardware impairment coefficient is shown in Fig. \ref{fig:7}. Here, $\Delta d$ denotes the distance by which the receiver moves away from its initial position. As $\kappa_t$ increases, the worst‑case SR decreases. When $\kappa_t$ becomes larger, the distortion caused by hardware impairments gradually intensifies, leading to a reduction in the final SR value. For directional antennas, when the user’s horizontal position remains unchanged, i.e., $\Delta d = 0$, $H_0 = 20$ m and $H_0 = 200$ m achieve comparable SR performance. This is because a height change of tens to hundreds of meters has a limited relative impact on the total propagation distance. When the height increases to $H_0 = 1000$ m, for example, when the antenna is deployed on a mountain top, the SR decreases noticeably. This is because an excessively high array height significantly increases the three‑dimensional propagation distance between the BS and the sea‑surface users, resulting in greater path loss. Therefore, increasing the antenna height does not necessarily improve the SR; its main advantage lies in extending the maritime visible range and improving propagation accessibility.
On the other hand, when the communication distance increases by $\Delta d = 1000$ m, the SR decreases due to greater path loss. Compared with isotropic antennas, directional antennas still exhibit a clear advantage under the same distance and hardware impairment conditions.

\begin{figure}
	\centering
	\includegraphics[width=0.47\textwidth, trim = 0 0 0 10,clip]{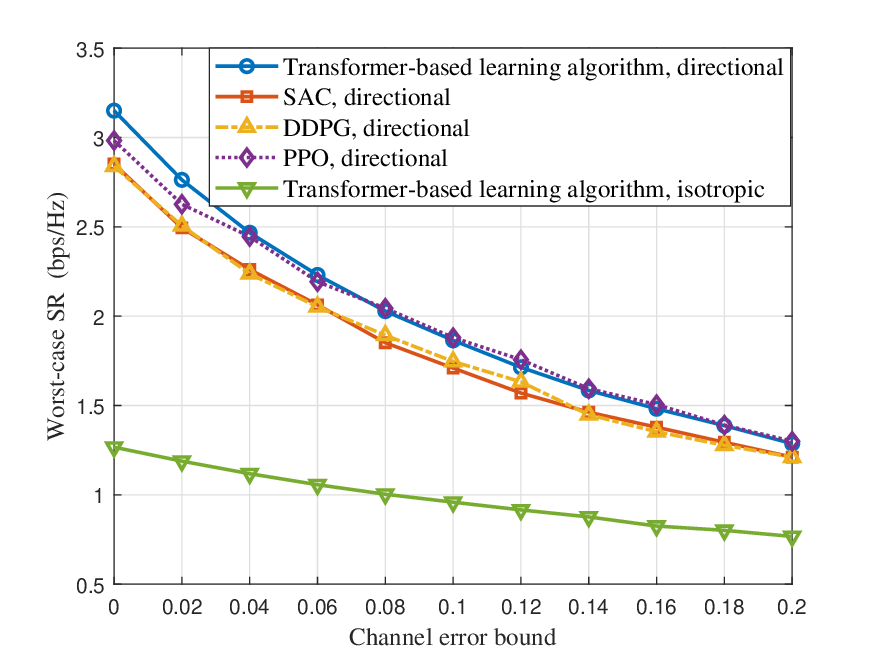}\\
	\caption{Illustration of  the worst-case SR versus channel error bound.}\label{fig:8}
\end{figure}

Fig. \ref{fig:8} compares the worst‑case SR of different methods under channel estimation errors. It can be observed that, as the channel error bound increases, the SR of all schemes decreases. This is because channel uncertainty causes a mismatch between the designed information beam and the true legitimate channel, and it also affects the AN suppression direction toward Eve as well as the zero‑interference condition at Bob.
It is worth noting that, over the entire error range, the schemes employing directional antennas outperform the Transformer‑based isotropic antenna scheme. This indicates that the proposed system is robust to imperfect channel state information (CSI).
\begin{figure}
	\centering
	\includegraphics[width=0.47\textwidth, trim = 0 0 0 10,clip]{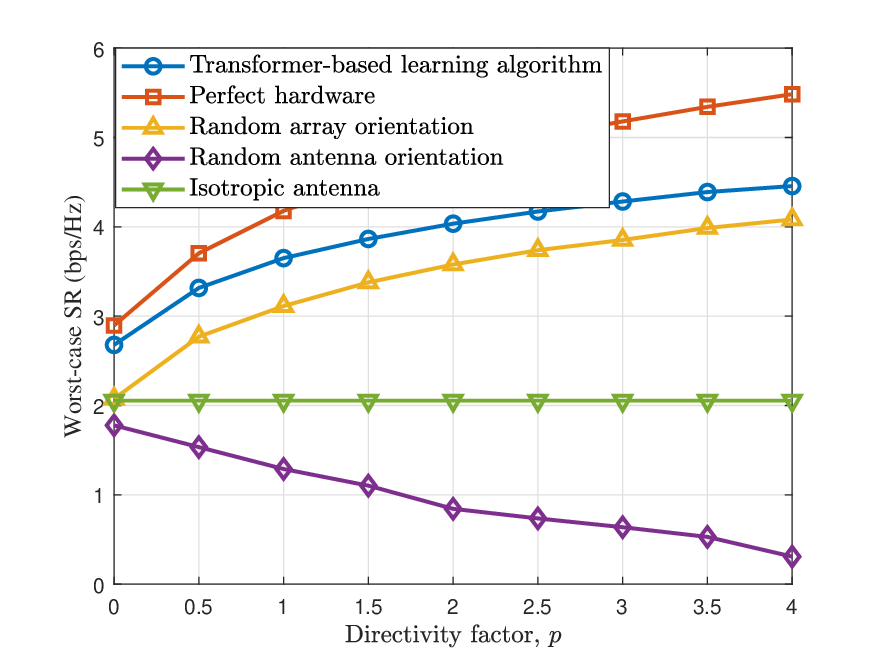}\\
	\caption{Illustration of  the worst-case SR versus directivity factor.}\label{fig:9}
\end{figure}

The worst‑case SR under different antenna directivity factors is shown in Fig. \ref{fig:9}. As $p$ increases, the main‑lobe directivity of the antenna is enhanced, and the beam becomes more concentrated. For the Transformer‑based learning algorithm and the perfect‑hardware scheme, since the antenna element orientations can be well controlled, the SR increases with $p$. Moreover, higher directivity can potentially enhance the effective gain toward Bob while reducing the energy leakage toward Eve. Although the random array orientation scheme can achieve a higher average SR as $p$ increases, its performance degrades significantly when the array deviates substantially from the user, and therefore it always performs worse than the Transformer scheme. In addition, the SR of the random antenna orientation scheme decreases as $p$ increases. If the antenna element orientations are randomly set, Bob is highly likely to fall outside the main lobe, and the gain loss caused by orientation mismatch becomes pronounced, thereby degrading the security performance.

The proposed antenna architecture can provide additional spatial degrees of freedom from three dimensions: array height, array pose, and antenna element orientation. Both the single‑user AO scheme and the multi‑user Transformer‑based learning algorithm can effectively exploit these degrees of freedom to achieve higher SR values. As the transmit power and the number of antennas increase, the worst‑case SR improves; however, as the number of users, hardware impairments, and CSI uncertainty increase, the security performance gradually degrades. Furthermore, stronger directivity can yield higher security performance gains only when combined with proper antenna orientation optimization. Taking a worst‑case SR of 5 bps/Hz as the reference, the proposed scheme can reduce the transmit power by approximately 10 dB compared with isotropic antennas, corresponding to about 
90\% transmit power savings. The required array size can be reduced from about 128 elements to about 16 elements, corresponding to approximately 87\% antenna savings. While guaranteeing a worst‑case SR of about 2 bps/Hz, the number of supported legitimate users can be increased from about 3 to about 5, an improvement of approximately 67\%.
\section{Conclusion}
This paper investigates elevation‑adjustable and rotatable secure communication in a maritime communication scenario. Convex optimization and learning‑based schemes are proposed for the single‑user and multi‑user cases, respectively. Simulation results validate the effectiveness of the proposed schemes. Under limited system resources, the security performance gain obtained by solely increasing the antenna height is limited due to path loss. By jointly optimizing the antenna height, array orientation, antenna orientation, and precoding vectors, a higher security performance gain can be achieved compared with conventional isotropic antenna setups. Moreover, the proposed Transformer‑based learning scheme outperforms conventional schemes. In maritime communication scenarios, elevation‑adjustable and rotatable antenna arrays can achieve higher security performance and robustness.

\ifCLASSOPTIONcaptionsoff
\newpage
\fi

\renewcommand\refname{References}
\bibliographystyle{IEEEtran}
\bibliography{mybib}

\newpage

\vfill

\end{document}